\documentclass[preprintnumbers, prd, onecolumn, floatfix, superscriptaddress, nofootinbib] {revtex4-2}
\usepackage{setspace}
\usepackage{graphicx}
\usepackage{subcaption}
\usepackage{epsfig}
\usepackage{bm}
\usepackage{amssymb}
\usepackage{float}
\usepackage{amsmath}
\usepackage{dcolumn}
\usepackage[colorlinks]{hyperref}
\usepackage[usenames,dvipsnames]{color}
\usepackage{enumitem}

\hypersetup{ breaklinks=true, pdfstartview={FitH}, colorlinks=true, linkcolor=blue, citecolor=red, filecolor=magenta, urlcolor=blue, anchorcolor=green, linktocpage=true }

\def\doi{http://doi.org}

\newcommand{\be}{\begin{equation}}
\newcommand{\ee}{\end{equation}}
\newcommand{\beano}{\begin{eqnarray*}}
\newcommand{\eeano}{\end{eqnarray*}}
\newcommand{\ba}{\begin{eqnarray}}
\newcommand{\ea}{\end{eqnarray}}

\begin{document}

\title{ Power law cosmology in Gauss-Bonnet gravity with pragmatic analysis}
\author{Rita Rani}
\email{rita.ma19@nsut.ac.in}
\affiliation{Department of Mathematics, Netaji Subhas University of Technology, New Delhi-110 078, India}
\author{Shaily}
\email{shailytyagi.iitkgp@gmail.com}  
\affiliation{School of Computer Science Engineering and Technology, Bennett University, Greater Noida, India}
\author{G. K. Goswami}
\email{gk.goswami9@gmail.com}
\affiliation{Department of Mathematics, Netaji Subhas University of Technology, New Delhi-110 078, India}
\author{J. K. Singh}
\email{jksingh@nsut.ac.in}
\affiliation{Department of Mathematics, Netaji Subhas University of Technology, New Delhi-110 078, India}

\begin{abstract}

\noindent
In this study, we present an approach $ f(R, G) $ gravity incorporating power law in $ G $. To study the cosmic evolution of the universe given by the reconstruction of the Hubble parameter given by $ E(z) = \bigg( 1+\frac{z(\alpha+(1+z)^{\beta})}{2 \beta + 1} \bigg)^{\frac{3}{2 \beta}} $. Subsequently, we use various recent observational datasets of OHD, Pantheon, and BAO to estimate the model parameters $ H_0,~\alpha $, and $ \beta $ applying the Markov Chain Monte Carlo (MCMC) technique in the emcee package to establish the validity of the model. In our findings, we observe that our model shows consistency with standard $ \Lambda $CDM, transits from deceleration to acceleration, and enters the quintessence region in late times. The cosmological model satisfies necessary energy constraints, simultaneously violating the strong energy condition (SEC), indicating a repulsive nature and consistent with accelerated expansion. The cosmic evolution of the Hawking temperature and the total entropy for the various observational datasets also show the validity of the model. Thus, our established model demonstrates sufficient potential for explicitly describing cosmological models.

\end{abstract}

\maketitle


\section{Introduction}
Recent observational modifications that have taken account of cosmological supernovae data  
\cite{SupernovaSearchTeam:1998fmf, SupernovaCosmologyProject:1998vns, Perlmutter:1999jt, WMAP:2003elm, WMAP:2006bqn, WMAP:2010qai, Planck:2018vyg} have indicated that the universe is expanding with acceleration. There are two explanations for this accelerated behavior of the universe. The first one is explained by the mysterious energy known as dark energy suggested by researchers that exerts high negative pressure. At present times \cite{Planck:2018vyg}, it contributes about $ 70\% $ of the total universe energy mass. Along with this, there is the existence of energy matter which is known as dark matter. In literature, various dark energy models have been discussed such as $ \Lambda $CDM model, Quintessence, Chaplygin gas, and Braneworld but the most preferred one is $ \Lambda $CDM \cite{Sahni:1999gb}. Several works support the dark energy theory; some are as  \cite{Chevallier:2000qy, Copeland:2006wr, amendola2010dark, Frieman:2008sn, Carroll:2000fy}. The second one is the modifications of the gravitational theory related to Einstein's field equations \cite{Nojiri:2006ri, Capozziello:2007ec, Lobo:2008sg}. General Theory of Relativity explains the modern cosmology. Various approaches have been developed over the years to describe the late-time cosmic acceleration such as modifying the energy-momentum tensor or the geometry in Einstein's field equations (EFEs). Recently, several authors have accomplished number of excellent works in the alternative theory of gravity \cite{deHaro:2023lbq, Singh:2018xjv, Aviles:2014rma, Capozziello:2019cav, delaCruz-Dombriz:2016bqh,  Shaily:2024nmy, Balhara:2023mgj, Shaily:2024xho, Singh:2024gtz, Pawar:2024juv, Shabani:2016dhj, Singh:2022nfm, Singh:2023bjx, Singh:2024gtz, Shaily:2024nmy, Singh:2023yye} 

The modified theories of gravity successfully described the acceleration in early and late-time eras in a unified way \cite{Nojiri:2003ft, Carroll:2003wy}. Most popular one is  $ f(R) $ gravity theory \cite{ Starobinsky:2007hu, Chiba:2006jp,  Bertolami:2007gv, Capozziello:2008qc, Sotiriou:2008rp, Nojiri:2010wj, Goswami:2022vfq} in which the Ricci scalar $ R $ is being replaced with an arbitrary function $ f(R) $ in the Einstein-Hilbert action of the general relativity, $ f(R, T) $ gravity where $ R $ and $ T $ denote the Ricci scalar and  energy momentum scalar respectively \cite{Harko:2011kv, Capozziello:2014bqa, Singh:2022jue, Pradhan:2023, Singh:2023gxd, Singh:2024ckh, Singh:2024kez}, $ f(T) $ gravity where $ T $ is the teleparallel gravity \cite{Ferraro:2006jd, Cai:2015emx, Paliathanasis:2016vsw, Duchaniya:2023aeu}, $ f(G) $ gravity in which $ G $ is the Gauss-Bonnet Tensor \cite{Townsend:1979js, Birrell_1982, Barth:1983hb, DeFelice:2008wz, MontelongoGarcia:2010ip, Sharif:2016drh, Boehmer:2021aji}, $ f(Q) $ gravity where $ Q $ is the non-metricity tensor \cite{Mandal:2020lyq, Mandal:2020buf, Frusciante:2021sio, Heisenberg:2023lru,  Goswami:2023knh}.

Among various modified theories, we have focused on the extension of $ f(R) $ and $ f(G) $ gravity that is $ f(R, G) $ gravity where arbitrary function depends on the Ricci scalar $ R $  and the Gauss-Bonnet invariant $ G $. The most widely used modified theory of gravity is $ f(R) $ gravity which depends on the arbitrary function of $ R $. Various alternative proposals to the $ f(R) $ gravity theory introduced one of them using Lovelock invariants. Several authors have successfully described the dark energy epoch in $ f(R, G) $ gravity theory \cite{ Nojiri:2005jg, Nojiri:2005am, Li:2007jm}. In the present paper, we work on one of the modified theories of gravity $ f(R, G) $. Here, the Gauss-Bonnet invariant involves a higher order of Ricci scalar $ R $, leading to some interesting cosmological effects. It explains both early and late times under the same geometrical criterion. The Gauss-Bonnet invariant $ (G) $ defined as $ G \equiv R^2 -4 R_{ij} R^{ij} + R_{ijkl} R^{ijkl} $  may arise naturally in the gauge theories like Chern-Simons \cite{Achucarro:1986uwr, Gomez:2011zzd, Raushan:2023pdv}, Love-lock theory  \cite{Lovelock:1971yv, Bajardi:2021hya}, or Born-Infeld gravity \cite{Leigh:1989jq, Tseytlin:1997csa}. The Gauss-Bonnet term $ G $ contributes to tracing the anomaly involved in higher-order curvature terms that do not vanish.   

Here, we discuss the dark energy scenarios in one of the modified theories of gravity $ f(R, G) $ where both $ R $ and $ G $ are in linear form which involves power law in $ G $. There are several studies conducted in a different form of arbitrary functions of $ f(R, G) $ in different scenarios such as Atazadeh and Darabi \cite{Atazadeh:2013cz} study the viability of models in $ f(R, G) $ gravity models using energy conditions. M De Laurentis et. al. \cite{DeLaurentis:2015fea} discusses the cosmological inflation in the $ f(R, G) $ gravity model framework. Odintsov et al. \cite{Odintsov:2018nch} explained the dynamics of inflation and dark energy scenarios in the $ f(R, G) $ gravity by proposing different arbitrary functional forms of $ f(R, G)$. Bhatti et. al. \cite{Bhatti:2020cjz} studied the length of shell, entropy, and energy in the aspect of gravastars in $ F(R, G) $ gravity. Lohakare et. al. \cite{Lohakare:2021yu o} study the cosmological model in $ f(R, G)$ gravity with time-varying deceleration parameters. Shekh \cite{Shekh:2021dsl} studies the dynamical analysis of anisotropic dark energy of cosmological models in $ f(R, G) $ gravity in anisotropic dark energy LRS Bianchi Type -I metric. Chanda et. al. \cite{Chanda:2023lfk} studied the late-time cosmology with exponential interactions in $ f(R, G) $ gravity. Shaily et. al. \cite{Shaily:2024rjq} studied the bouncing cosmology in $ f(R, G) $ gravity with thermodynamical analysis. Das et. al. \cite{Das:2023bff} investigated the stability of spherically symmetric relativistic stellar configuration in modified gravity $ f(R, G) $ gravity. Various cosmological models have been studied in the $ f(R, G) $ gravity \cite{Bamba:2010wfw, DeFelice:2011ka, Odintsov:2015uca, Wu:2015maa, Lansberg:2016deg, Calza:2019egu, Elizalde:2020zcb, Mustafa:2020jln, Nada:2020jay, Naz:2023pfl, Singh:2022gln}.

This paper is set up as follows: The basic introduction of $ f(R, G) $ gravity has been discussed in Sec. \ref{sec-2}. In Sec. \ref{sec-3}, we have calculated the field equations by considering the arbitrary non-minimal coupled form of $ f(R, G) $ as $ f(R) + f(G) $. Also, we have calculated the physical cosmological parameters to analyze the universe's evolution. Further, we choose an appropriate ansatz of the Hubble parameter to solve the field equations. In Sec. \ref{sec-4}, we have discussed statistical methods to figure out the best-fit values of different free model parameters. Also, we plotted the error plot to compare our model with the standard $\Lambda$CDM model. Moreover, we discuss the deceleration, jerk parameters, statefinder, and $ Om $ diagnostics in the next section. In Sec. \ref{sec-6}, we observe the Energy Conditions for the model. In Sec \ref{sec-7}, we observe the behavior of Hawking's temperature and total entropy. In the last section \ref{sec-8}, the work is concluded with the finding.  


\section{$ f(R, G) $ gravity }{\label{sec-2}}

We consider a general formalism of $ f(R, G) $ gravity where gravitational field equations in $ f(R, G) $ gravity are derived from the following given action, 

\begin{equation}{\label{1}}
    S = \int\Big(\frac{1}{16\pi G_N} f(R,G) + L_m\Big)\sqrt{-g} dx^4,
\end{equation}

where $ L_m $ stands for the matter Lagrangian, $ G_N $ denotes the Newtonian constant and $ f(R,G) $ is an arbitrary function of Ricci Scalar ($ R $) and  Gauss-Bonnet invariant ($ G $) which is defined as,

\begin{equation}{\label{2}}
      G=R^2 -4 R_{ij} R^{ij} + R_{ijkl} R^{ijkl},
\end{equation}

where $ R_{ij}$ and $ R_{ijkl} $ indicate the Ricci tensor the Reimann tensor. Now, by varying the action in Eq. (\ref{2}) \textit{w.r.to} the metric tensor $ g_{ij} $, one can obtain the field equations as \cite{Odintsov:2015uca}

\begin{align*}{\label{3}}
     G_{ij} =& \frac{1}{f_R} \bigg[  8 \pi G_N T_{ij} + \frac{1}{2} g_{ij} \bigg( f(R,G)-R f_R \bigg)  +\nabla_i \nabla_j f_R - g_{ij} \Box  f_R  \\
    &+ f_G \bigg( -2 R R_{ij} + 4 R_{ik} {R^k}_{j} -2 {R^{klm}}_i R_{jklm} + 4 g^{kl} g^{mn} R_{ikjm} R_{ln} \bigg) \\
    &+2(\nabla_i \nabla_j f_G) R -2 g_{ij} (\Box f_G) R +4(\Box f_G) R_{ij} - 4 (\nabla_i \nabla_j f_G) {R^k}_j  \\
    &-4(\nabla_k \nabla_j f_G) {R^k}_i + 4 g_{ij} (\nabla_k \nabla_l f_G) R^{kl} -4 (\nabla_k \nabla_l f_G) g^{mk} g^{nl} R_{imjn} \bigg], 
\end{align*}

where $ G_{ij} $, $ T_{ij} $, and $ \nabla_i $ denote Einstein tensor, energy-momentum tensor, and covariant derivative operator associated with $ g_{ij} $ respectively. Here, the d'Alembert operator is defined as $ \Box \equiv g^{ij} \nabla_i \nabla_j $. Also, the notations $ f_R $ and $ f_G $ denote the partial derivatives of $ f $ concerning $ R $ and $ G $, respectively.


\section{Metric and Field Equations} {\label{sec-3}}

Let us consider the spatially flat FLRW metric,
 \begin{equation}{\label{4}}
      ds^{2} = -dt^{2} + a^{2}(t)(dx^{2}+dy^{2}+dz^{2}),
 \end{equation}
where $ a(t) $ denotes the scale factor of the universe. For flat FLRW metric, the expression of Ricci scalar $ R $ and Gauss-Bonnet term $ G $ in the form of Hubble parameter can be obtained as

\begin{equation}{\label{5}}
     R = 6 (2H^2 +\dot{H}),
\end{equation}

\begin{equation}{\label{6}}
     G = 24 H^2 (H^2 + \dot{H}).
\end{equation}

Also, the tensor of energy-momentum of the perfect fluid is $ T_{ij}= (\rho +p )u_i u_j + p g_{ij} $ and on considering $ f(R, G) = R + f(G) $, the field Equation can be written as,
\begin{align*}
    R_{ij} - \frac{1}{2} g_{ij} f(G)   =&   8 \pi G_N T_{ij} +\nabla_i \nabla_j f_R - g_{ij} \Box  f_R  \\
    &+ f_G \bigg( -2 R R_{ij} + 4 R_{ik} {R^k}_{j} -2 {R^{klm}}_i R_{jklm} + 4 g^{kl} g^{mn} R_{ikjm} R_{ln} \bigg) \\
    &+2(\nabla_i \nabla_j f_G) R -2 g_{ij} (\Box f_G) R +4(\Box f_G) R_{ij} - 4 (\nabla_i \nabla_j f_G) {R^k}_j      \\
    &-4(\nabla_k \nabla_j f_G) {R^k}_i + 4 g_{ij} (\nabla_k \nabla_l f_G) R^{kl} -4 (\nabla_k \nabla_l f_G) g^{mk} g^{nl} R_{imjn}.
    {\label{8}}
\end{align*}

We have assumed $ 8 \pi G_N = 2 \kappa $ for the easy interpretation. In the case of FLRW metric (\ref{4}), the Friedmann equations are,

\begin{equation}{\label{9}}
     6 H^2 + f(G) -G f_G +24 H^3 \dot{G} f_{GG} = 2  \kappa  \rho,
\end{equation}

\begin{equation}{\label{10}}
     4 \dot{H} +6 H^2 + f(G) - G f_G + 16 H \dot{G} \bigg( \dot{H} +H^2 \bigg) f_{GG} + 8 H^2 \ddot{G} f_{GG} + 8 H^2 \dot{G}^2 f_{GGG} =-2  \kappa  p,
\end{equation}
where $ \rho $ and $ p $ are the energy density and isotropic matter pressure, respectively.

In this work, we assume the function $ f(R, G) = R + f(G) $  where $ f(G) =  G^{1-m} $ with positive values of arbitrary constant $m$. It is observed that corresponding to a constant value of $ m $ that is $ m=1 $, we obtain $ f(R) $ gravity model. We took the inspiration of taking power form for $ G $ from  \cite{Shekh:2021dsl} to analyze our results. Therefore, the Friedman field equations are obtained after substituting $ f(G) =  G^{1-m} $. 

To analyze the theoretical results with the cosmological observations, we consider the independent variable redshift $ z $, instead of the usual time $ t $, it is defined according to 
\begin{equation}{\label{11}}
     1+z = \frac{a_0}{a} = \frac{1}{a},
\end{equation}
where we have imposed the condition that $ a_0=a(0) =1 $ and calculate the derivatives with respect to the time from the derivatives with respect to the redshift as
\begin{equation}\label{12}
      \frac{d}{dt} = \frac{dz}{dt} \frac{d}{dz} = -(1+z) H(z) \frac{d}{dz}
\end{equation}

The above field equations in terms of redshift can be written as 
\begin{align*}
    \rho (z) =& \frac{1}{ \kappa } \bigg( 3^{1-m} 8^{-m} H(z)^5 (H(z)^3 (H(z)-(1 + z) H'(z)))^{-m-1} \bigg(\frac{24^m (H(z)^3 (H(z)-(1+z) H'(z)))^{m+1}}{H(z)^3}\\
     & + 4 m H(z) ((3 m-2) (1+z)^2 H'(z)^2 + (1+z) H(z) ((m-1) (1+z) H''(z)+(1-3 m) H'(z)) + H(z)^2) \bigg) \bigg),
\end{align*}

\begin{align*}
    p(z)=& -\frac{1}{2  \kappa } \bigg( 24^{1-m} (H(z)^3 (H(z)-(1+z) H'(z)))^{1-m}+24^{1-m} (m-1) (H(z)^3 (H(z)-(1+z) H'(z)))^{1-m} \\ 
    &-4 (1+z) H(z) H'(z) + 2^{4-3 m} 3^{-m} (m-1) m (1+z) H(z)^2 (H(z)^3 (H(z)-(1+z) H'(z)))^{-m} (3 (1+z) H'(z)^2 \\
    &+ H(z) ((1+z) H''(z)-3 H'(z)))-\frac{1}{(1+z) H'(z)-H(z)} \bigg( 3^{-m} 8^{1-m} (m-1) m (1+z) H(z)^2 (H(z)^3 (H(z) \\
    &-(1+z) H'(z)))^{-m}((m+1) (1+z) (3 (1+z) H'(z)^2 + H(z) ((1+z) H''(z)-3 H'(z)))^2 + (H(z)-(1+z) H'(z))^2 \\
    &(9 (1+z)^2 H'(z)^3+ 2 (1+z) H(z) H'(z) (5 (1+z) H''(z)-3 H'(z)) + H(z)^2 ((1+z) ((1+z) H'''(z) - H''(z))\\
    &-3 H'(z))))+ 6 H(z)^2 \bigg).
\end{align*}
The evaluated field equations have three unknowns $ \rho $, $ p $, and $ H $. To solve the equations, we need to consider the ansatz of one of the unknowns. Here, we consider the ansatz of Hubble parameter $ H $ in terms of redshift ($ z $) to solve the above equations. The motivation behind this ansatz of a particular form is from the work of Myrzakulov et. al. \cite{Myrzakulov:2023sir}. The ansatz of the Hubble parameter is, 
 
\begin{equation}{\label{14}}
    E(z) = \bigg( 1+\frac{z(\alpha+(1+z)^{\beta})}{2 \beta + 1} \bigg)^{\frac{3}{2 \beta}}
\end{equation}
where $ E(z)=\frac{H(z)}{H_0 }$, $ H_0  $ is the Hubble parameter at present and $ \alpha $, $ \beta $ are the model parameters. To constrain these parameters, we apply the MCMC technique in the emcee package and analyze the $ f(R, G) $ function.


\section{Observational Analysis}\label{sec-4}

In this section, we use some different observational datasets to approximate the parameters $ H_0 $, $ \alpha $ and $ \beta $. We employ the conventional Bayesian methodology to analyze the observational data and also the Markov Chain Monte Carlo (MCMC) technique to derive the posterior distributions of the parameters \cite{Foreman-Mackey:2012any}. Here, we use different datasets (i) the Hubble dataset (77 points), (ii) the Pantheon data (1048 data points), (iii) the OHD + Pantheon data, and (iv) the OHD + Pantheon + BAO dataset. 

\subsection{The Hubble dataset}
The Hubble data, which relies on the varying ages of galaxies and their connection to the universe's expansion history, is a valuable instrument for comprehending the presence of Dark Energy and Dark Matter in the cosmos. The Hubble data relies on the varying ages of the galaxies and is related to the history of the universe's expansion. Thus, it is a handy tool for understanding the Dark Energy and Dark matter present in the universe. The Hubble parameter explains the rate of expansion. In this work, we use $ H(z) $ dataset with $ 77 $ data points \cite{Shaily:2022enj} for different redshift within the range ($ 0 \leq z \leq 2.36 $). Now, we use the Chi-square function to estimate the parameters $ H_0 $, $ \alpha $, and $ \beta $ and it is defined as
\begin{equation} \label{chi}
   {\chi^{2}}_{OHD}{(H_0, \alpha, \beta)}=\sum\limits_{i=1}^{77}\frac{[H(H_0, \alpha, \beta, z_{i})-H_{obs}(z_{i})]^{2}}{ \sigma^2 (z_{i})},
\end{equation}
where $ H(H_0, \alpha, \beta)$ and $ H_{obs} $ denote theoretical, and observed values of Hubble parameter respectively, and $ \sigma{(z_{i})} $ denotes the standard error in Hubble data set.\\

\begin{figure}
    \begin{center}
        \subfloat[]{\label{HubMCMC} \includegraphics[scale=0.75]{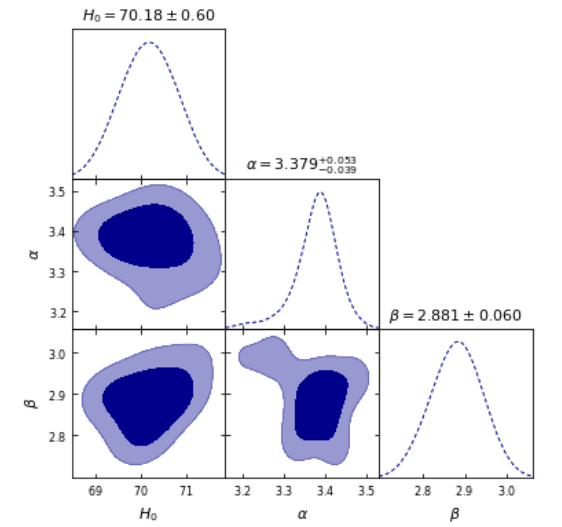}}
        \subfloat[]{\label{UnionMCMC} \includegraphics[scale=0.62]{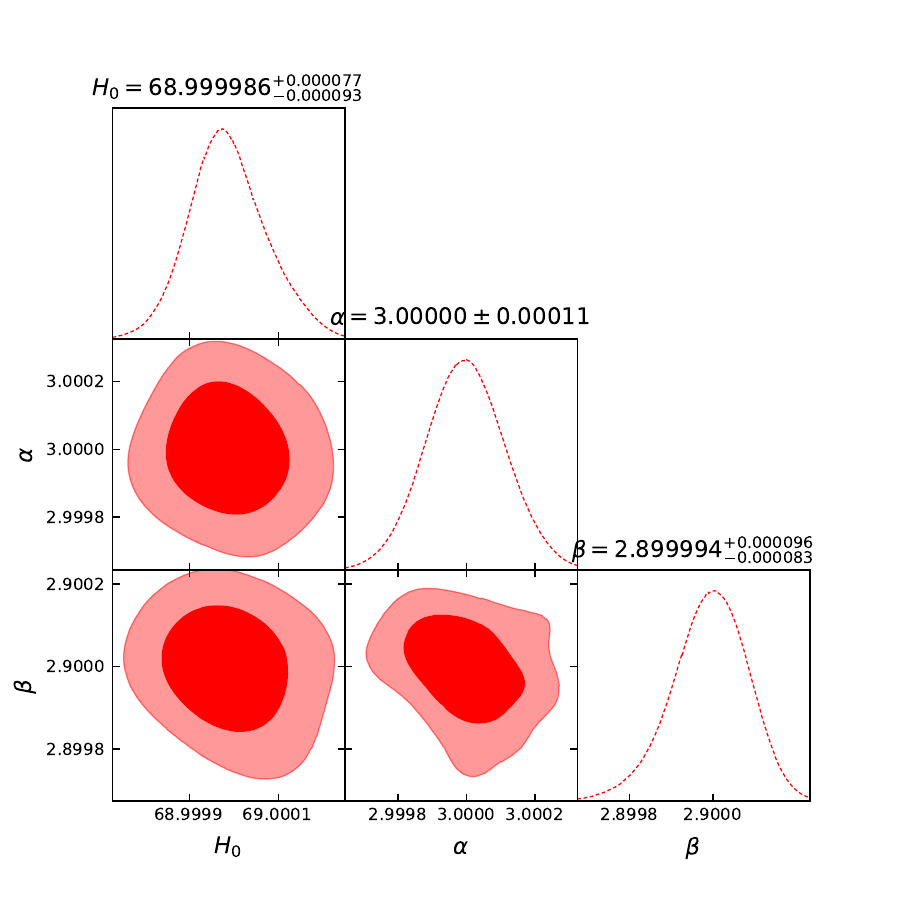}}\\
         \subfloat[]{\label{PanBaoMCMC} \includegraphics[scale=0.62]{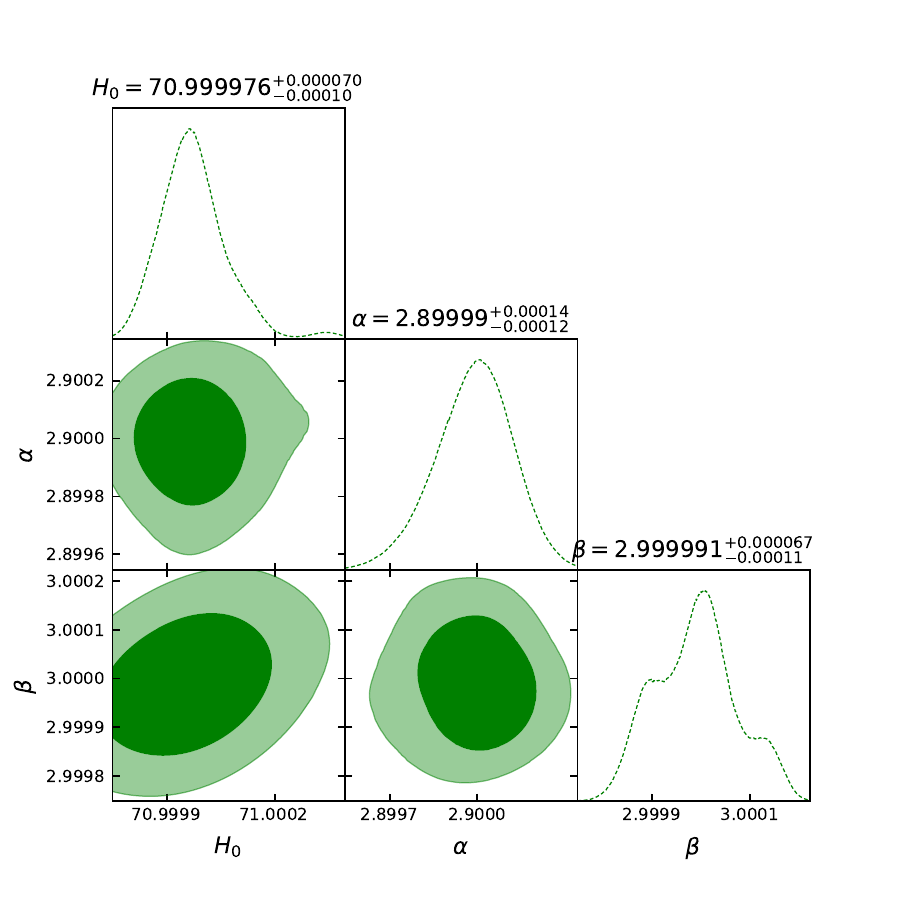}}
        \subfloat[]{\label{HzPAnBaoMCMC} \includegraphics[scale=0.62]{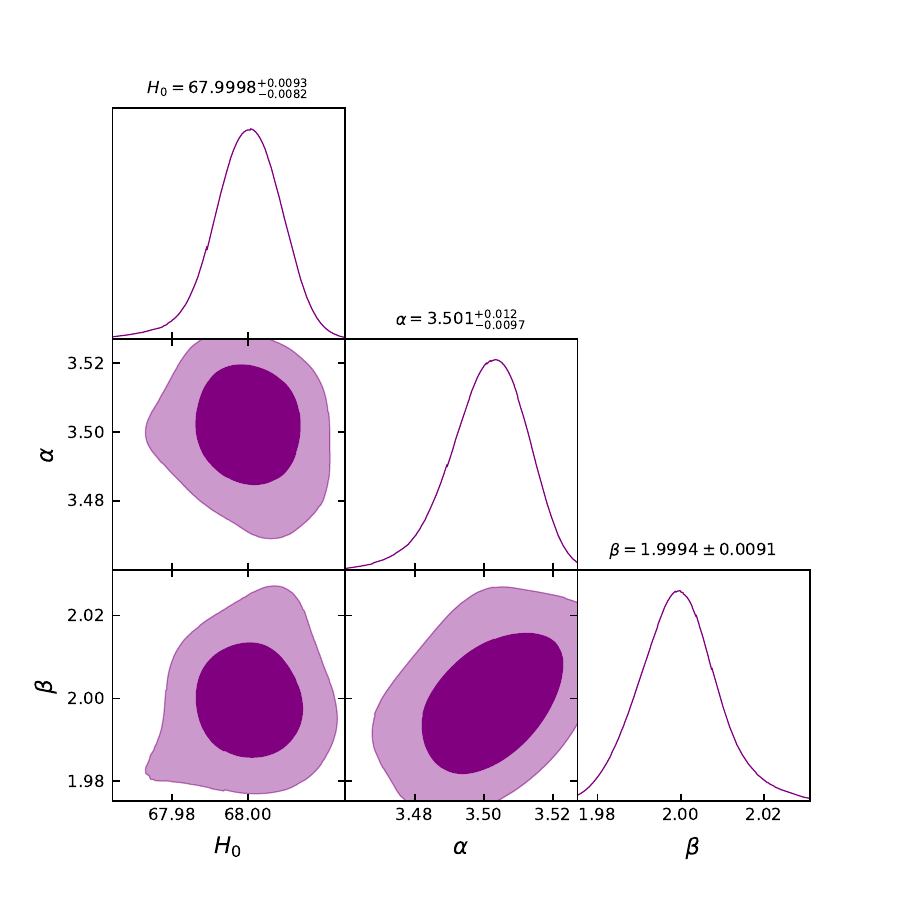}}
     \end{center}
    \caption{{\it The likelihood contours, with $ 1\sigma $ and $ 2\sigma $ confidence levels (CLs), for $ H(z) $  data (upper left),  $ Pantheon $ data (upper right), for $ Pantheon+BAO $ data (lower left), and the joint analysis of  $ H(z)+ Pantheon + BAO $ data (lower right). 
}}
\label{Hz}
\end{figure}

\subsection{Pantheon Data}
We have used the Pantheon dataset compilation to find the optimized values of the model parameters $ H_0 $, $ \alpha $, and $ \beta $.

\begin{equation} 
    {\chi^2}_{Pan}(\alpha,\beta,H0) = \sum\limits_{i=1}^{1048}\frac{[\mu_{th}(\alpha,\beta,H0,z_{i})-\mu_{obs}(z_{i})]^{2}}{err_{\mu} {(z_{i})}^{2}},
\end{equation}

where $ \mu_{th} $ and $ \mu_{obs} $ depict the theoretical and observed distance modulus of the model. $ err_{\mu}$ is the standard error in that. 
The distance modulus $ \mu(z) $ can be defined as 
\begin{equation}
      \mu(z) = m- M = 5 \log D_l(z) + \mu_0,
\end{equation}
where $ m $ and $ M $ represent a standard source's apparent and absolute magnitude respectively.
The Luminosity distance is defined as \cite{Copeland:2006wr},
\begin{equation}
   D_l(z) = c (1+z)  \int_{0}^{z} \frac{dz*}{H(z*)},
\end{equation}
 and nuisance parameter $ \mu_0 $
 \begin{equation}
     \mu_0 = 25+ 5 \log(\frac{{H_0}^{-1}}{1Mpc} ).
 \end{equation}

\begin{figure}
    \begin{center}
        \subfloat[]{\label{Hzerr} \includegraphics[scale=0.37]{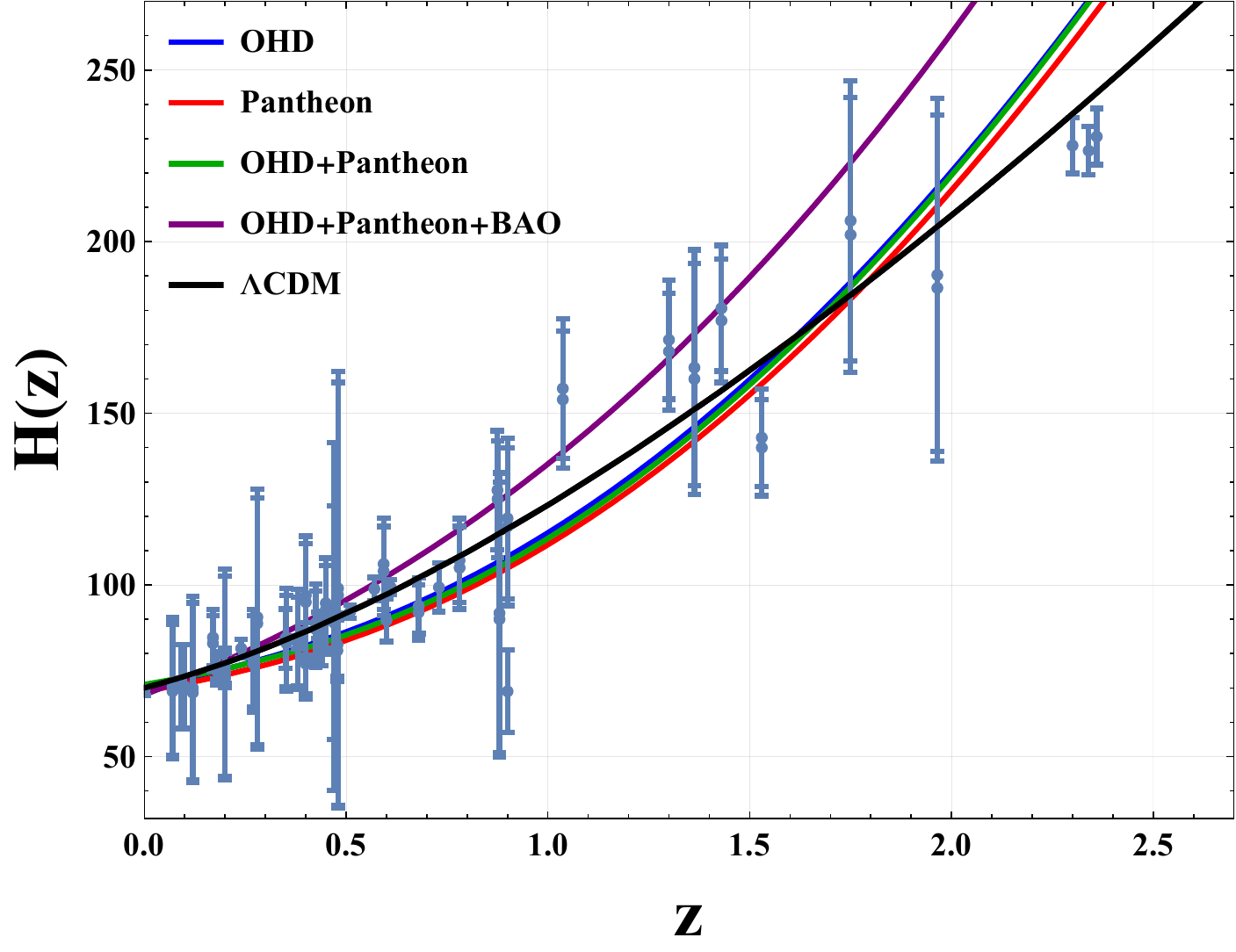}}\hfill
        \subfloat[]{\label{disterr} \includegraphics[scale=0.44]{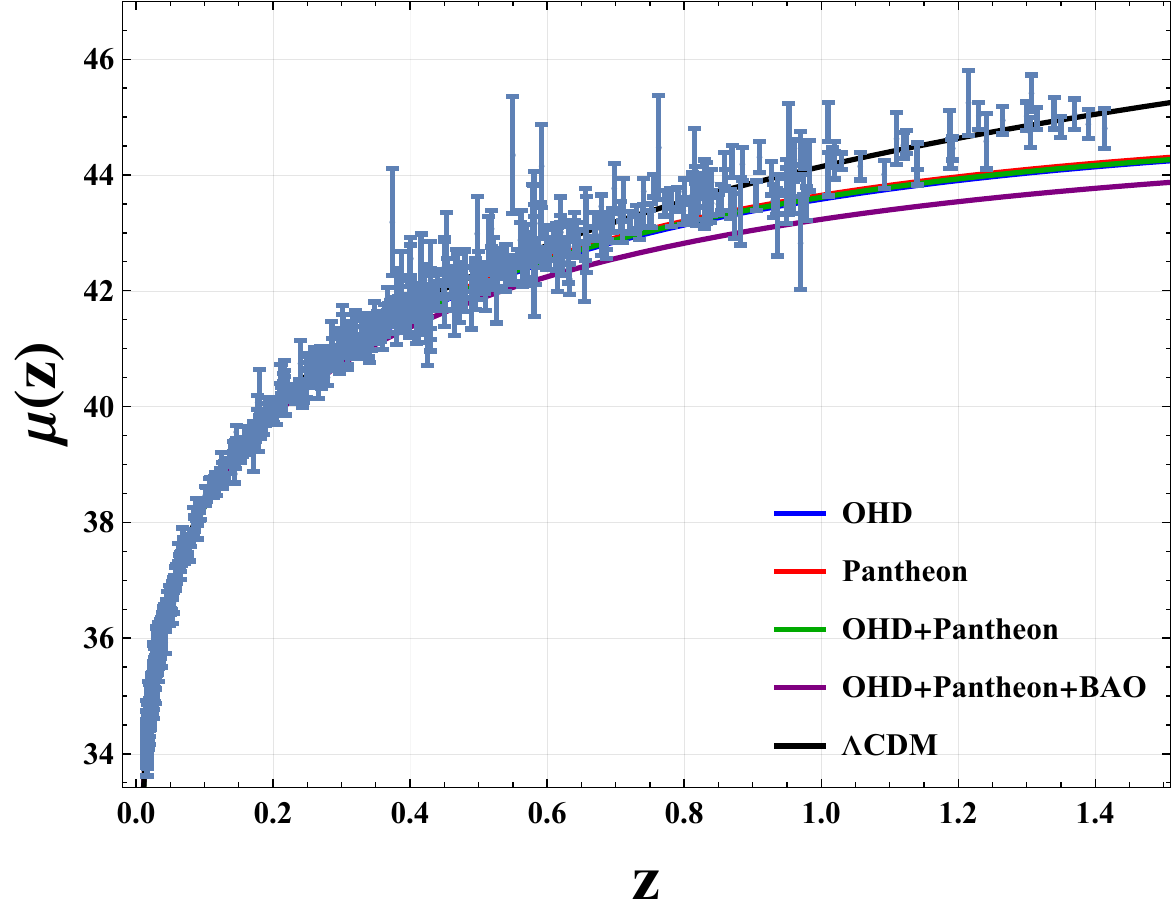}}\par
        \subfloat[]{\label{mb} \includegraphics[scale=0.44]{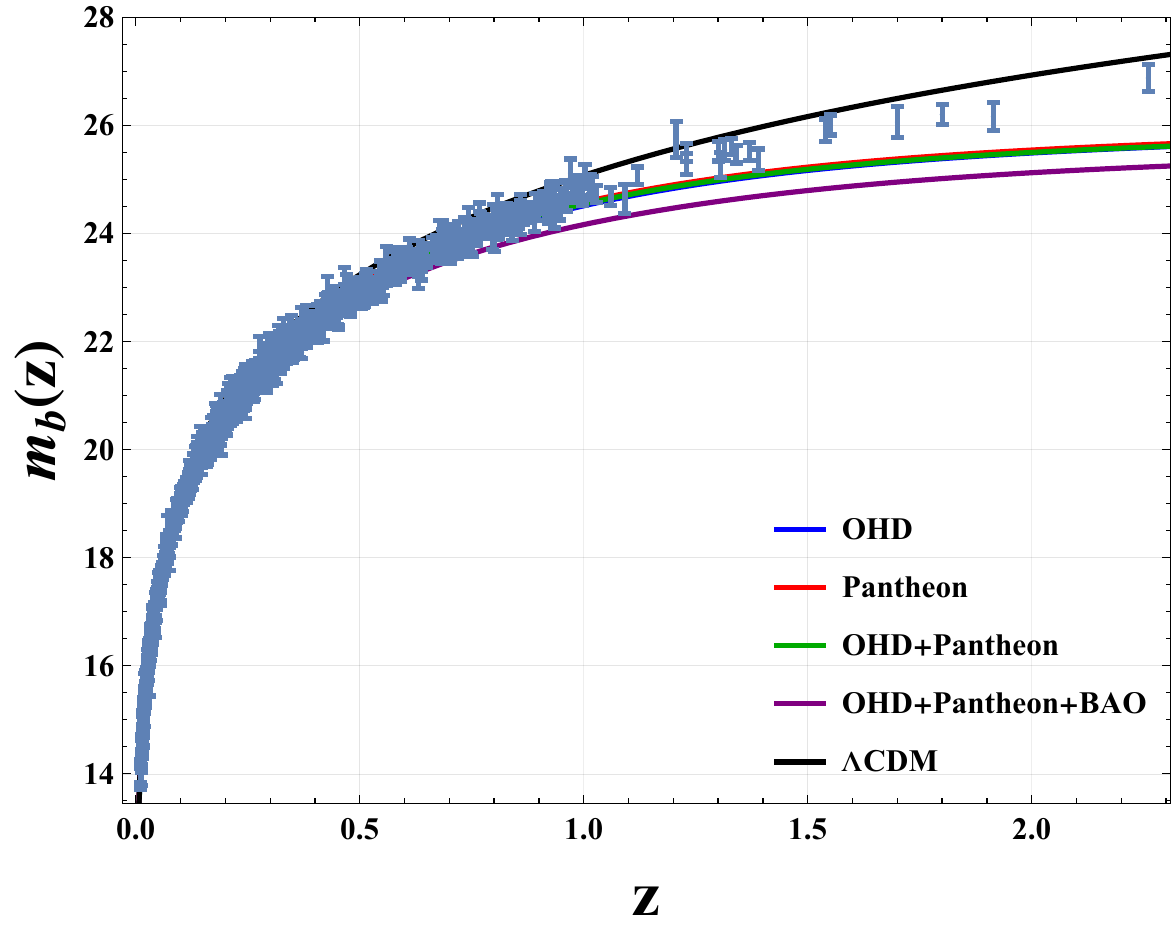}}
    \end{center}
\caption{{\it The error bar plots for the Hubble parameter $ H(z) $, distance modulus $ \mu(z) $, and the apparent magnitudes $ m_b(z) $ using 77 data points of OHD, 580 data points of SNIa and 1048 Pantheon data respectively for four different observed values of 3-tuple $ H_0 $, $ \alpha $, and $ \beta $. These plots exhibit the degree of consistency between the presented model and the $ \Lambda $CDM.
}}   
\label{err} 
\end{figure}

\subsection{ BAO Datasets}
BAO studies the oscillations in the early Universe generated by cosmic perturbations in a fluid composed of photons, baryons, and dark matter and connected via Thomson scattering. In this work, we consider a sample of BAO distance measurements from different surveys such as SDSS(R) \cite{Padmanabhan:2012hf}, the 6d F Galaxy survey \cite{Beutler:2011hx}, BOSS CMASS \cite{BOSS:2013rlg}, and three parallel measurements from the Wiggle Z survey \cite{Blake:2012pj, Blake:2011en, SDSS:2009ocz, WMAP:2012nax}.
To obtain analysis based on BAO the following equations are used:
The distance redshift ratio is given by 
\begin{equation}
    d_z= \frac{r_s(z*)}{D_v(z)},
\end{equation}
where $ r_s(z*) $ denoted the co-moving sound horizon at the time when photons decouple \cite{SDSS:2005xqv}, defined as  $r_s(a) = \int_0^a \frac{c_s da}{ a^2 H(a) } $ and $ z* $ is the photons decoupling redshift. Here we consider $ z* = 1090 $ for the analysis.
The dilation scale is denoted by $ D_v(z)$ which is denoted by 
\begin{equation}
    D_v(z) = \bigg(\frac{z {d_A}^2 (z)}{ H(z)} \bigg)^{\frac{1}{3}}, 
\end{equation}
where $ d_A(z) $ represents the angular diameter distance which is defined as $  d_A(z*) = c \int_0^{z*} \frac{dz'}{H(z')}  $.

The $ \chi^{2}_{BAO} $ corresponding to BAO measurements is given by \cite{giostri2012cosmic}
\begin{equation}
    \chi_{BAO}^2 = X^{T} C^{-1} X.
\end{equation}

\begin{equation}
X = 
\begin{bmatrix}
$$ \frac{d_A(z*)}{D_v(0.106)}-30.84$$ \\
$$ \frac{d_A(z*)}{D_v(0.35)}- 10.33 $$\\
$$ \frac{d_A(z*)}{D_v(0.57)}- 6.72 $$\\
$$ \frac{d_A(z*)}{D_v(0.44)}- 8.41 $$\\
$$ \frac{d_A(z*)}{D_v(0.6)}- 6.66 $$\\
$$ \frac{d_A(z*)}{D_v(0.73)}- 5.43 $$
\end{bmatrix}
\end{equation}

and $ C^{-1} $ is the inverse of the covariance matrix given by \cite{giostri2012cosmic}

\begin{equation}
C^{-1} = 
\begin{bmatrix}
 0.52552 & -0.03548 & -0.07733 & -0.00167 & -0.00532 & -0.0059 \\
 -0.03548 & 24.9707 & -1.25461 & -0.02704 & -0.08633 & -0.09579 \\
 -0.07733 & -1.25461 & 82.9295 & -0.05895 & -0.18819 & -0.20881 \\
 -0.00167 & -0.02704 & -0.05895 & 2.9115 & -2.98873 & 1.43206 \\
 -0.00532 & -0.08633 & -0.18819 & -2.98873 & 15.9683 & -7.70636 \\
 -0.0059 & -0.09579 & -0.20881 & 1.43206 & -7.70636 & 15.2814 \\
\end{bmatrix}
\end{equation}

\subsection{Joint dataset}
We use a combination of the Hubble, Pantheon, and BAO datasets to estimate the parameters through their collective likelihood function.
The $ \chi^2_{OHD+PAN+BAO} $ function for the combined datasets is given by,
\begin{equation}
    \chi^2_{OHD+PAN+BAO} = \chi^2_{OHD} + \chi^2_{PAN}+ \chi^2_{BAO}
\end{equation}

Fig. \ref{Hzerr} and \ref{disterr} represent the error bar plots of the model with Hubble datasets and Pantheon datasets respectively. It is observed that the models depict similar behavior as the $ \Lambda $CDM model in both figures.

\section{Cosmological Parameters of the Model}{\label{sec-5}}
\subsection{Deceleration Parameter (DP)}
The deceleration parameter $ (q) $ helps us understand the dynamics of the universe. The phase transition from deceleration to acceleration explains the expansion of the universe. The relation with the Hubble parameter can calculate the deceleration parameter in terms of redshift $ z $ \cite{Visser:2003vq, Liu:2023agr},
\begin{equation}{\label{15}}
      q(z) = -1 + (1+z) \frac{H'(z)}{H(z)}.
\end{equation}
Using Eq. (\ref{14}) in Eq. (\ref{15}), we get
\begin{equation}{\label{16}}
     q(z) = -1 + \frac{3 \left(\alpha (1+z) + (1+z)^{\beta } ( 1+ z + z \beta) \right)}{2 \beta (1 + z ((1 + z)^\beta + \alpha) + 2 \beta ) }.
\end{equation}

\begin{figure}[h]
    \begin{center}
        \subfloat[]{\label{qz} \includegraphics[scale=0.55]{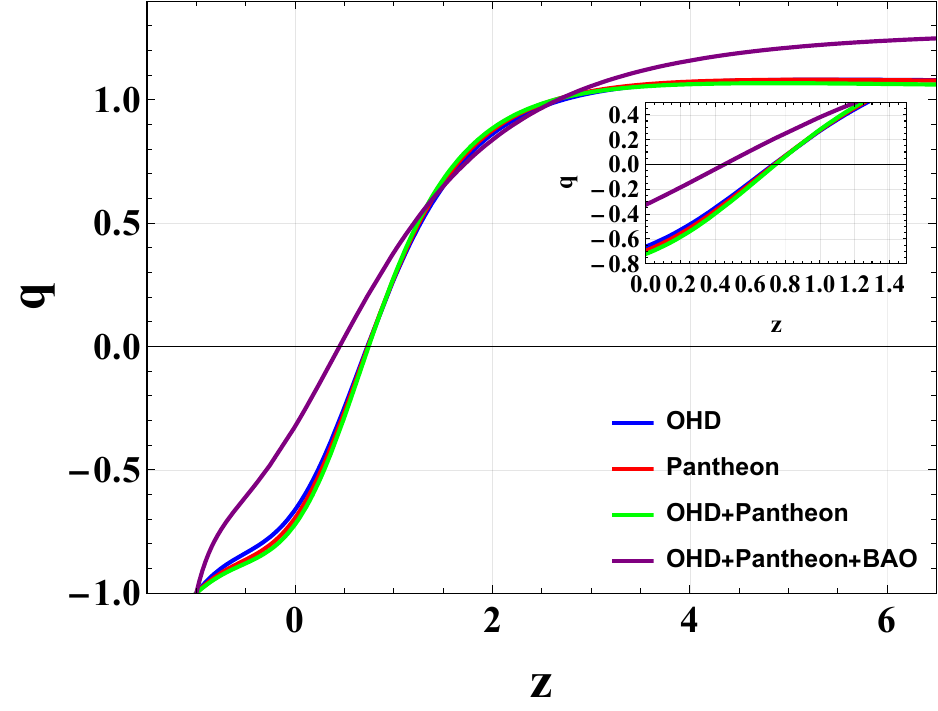}}
        \subfloat[]{\label{jz} \includegraphics[scale=0.52]{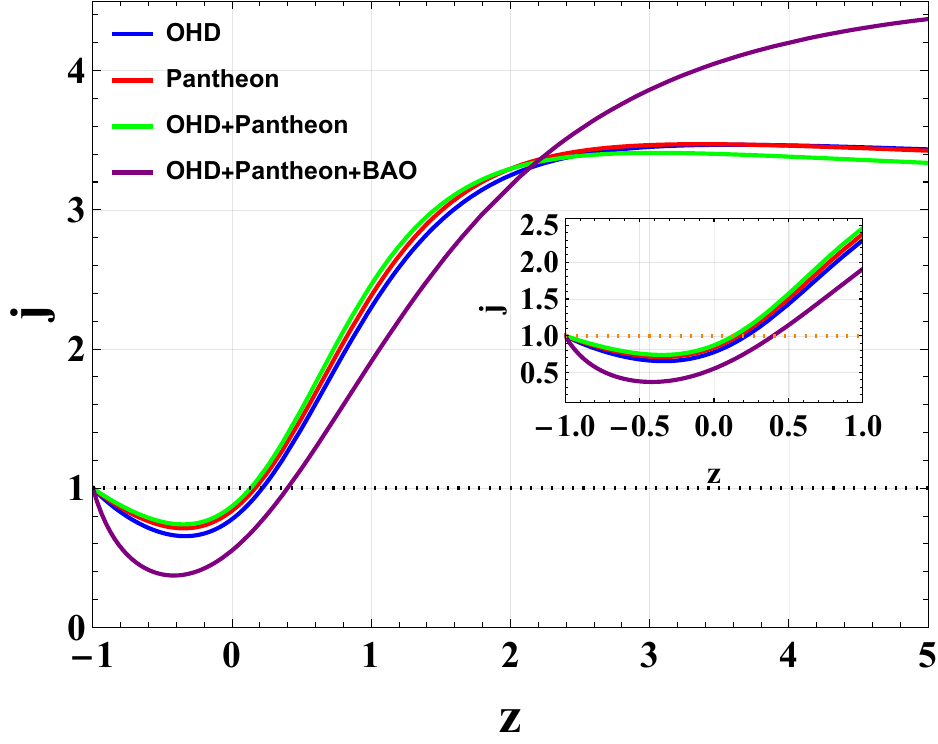}}\\
         \subfloat[]{\label{pz} \includegraphics[scale=0.55]{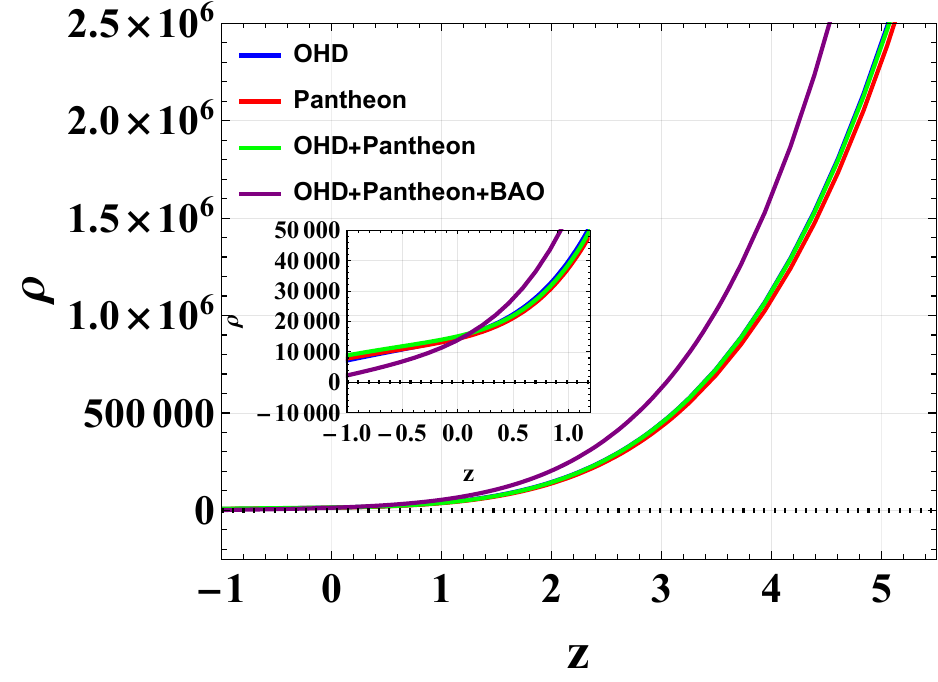}}
        \subfloat[]{\label{rhoz} \includegraphics[scale=0.55]{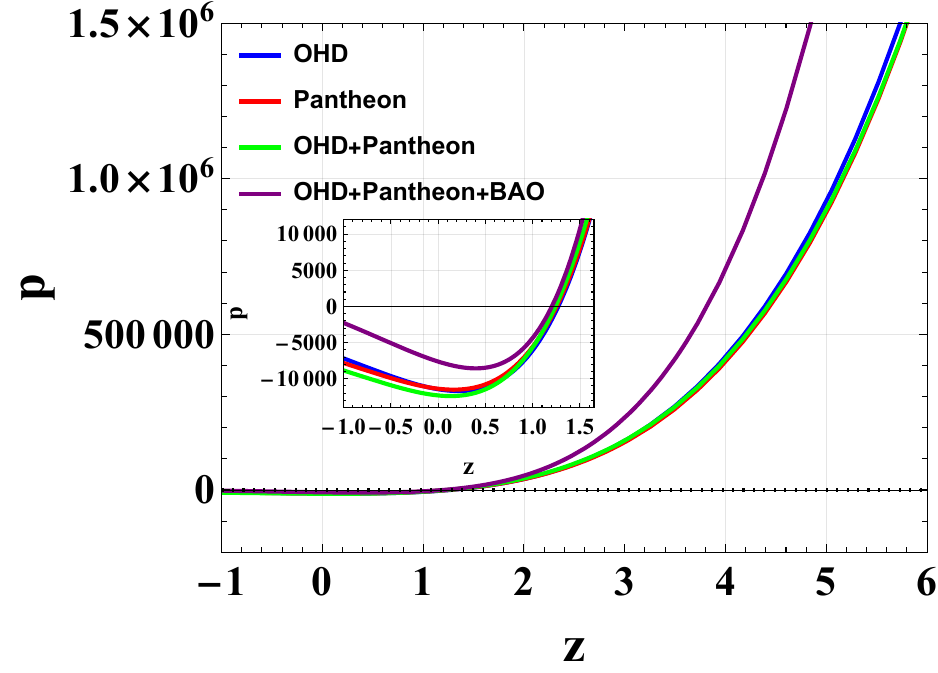}}
    \end{center}
    \caption{{\it The evolution of the deceleration parameter $ q(z) $, Jerk parameter $ j(z) $, density parameter $ \rho (z) $, and the isotropic pressure $ p (z) $ for the different constrained values of the model parameters, and the arbitrary constants $ m= 5.62 $, and $ \kappa \simeq 1 $.}}
\label{qz}
\end{figure}

In Fig. (\ref{qz}), it is observed that there is a phase transition from the decelerating to the accelerating phase. The observed phase transition is $ z_{tr}(OHD) \equiv 0.734 $, $ z_{tr}(PAN) \equiv 0.739 $, $ z_{tr}(OHD+PAN) \equiv 0.746 $ and $ z_{tr}(OHD+PAN+BAO) \equiv 0.45 $ for two different sets of values of model parameters as obtained from Hubble and Pantheon datasets respectively. The evaluated values of the deceleration parameter at $ z=0 $ that is $ (q_0) $ lie within the limits over the cosmographic coefficients in the recent analysis \cite{Capozziello:2014zda, Capozziello:2015rda}

\begin{figure}[tbph]
\begin{center}
   \includegraphics[scale=0.55]{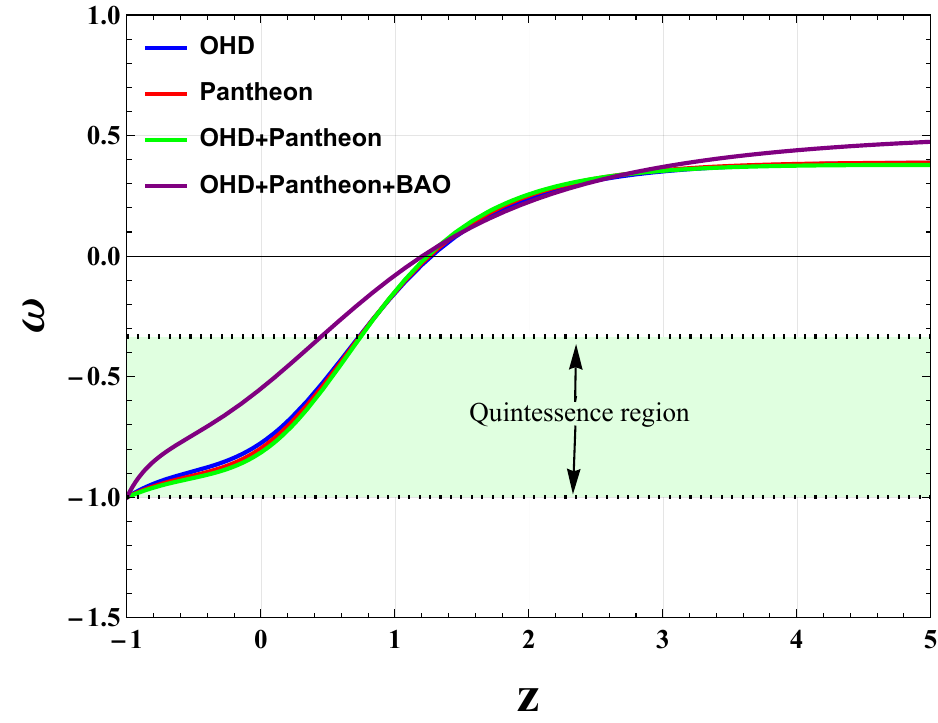}
\end{center}
\caption{{\it The evolution of EoS parameter $ \omega (z) $ for the different constrained values of the model parameters, and the arbitrary constants $ m= 5.62 $, and $ \kappa \simeq 1 $.}}
\label{wz}
\end{figure}

\begin{table}
\caption{ Best-fit values of the model parameters and present value of DP with phase transition point }
\label{tab: parameters value} 
\begin{center}
\begin{tabular}{l c c c c c} 
\hline\hline
     \\ 
    Parameters~~~~~~~~~~~~~~~~& ~~~~ OHD ~~~~~~~~~~~~ &~~~~~~~ Pantheon data ~~~~~~~~& ~~~~ OHD+Pan  ~~~~~~~~~~~~ &~~~~~~~ OHD+Pan+BAO
        \\
        \hline
        \\  
       $ H_0 $ & $ 70.18 \pm 0.60 $ & $ 68.99^{+0.000077}_{-0.000093} $ & $ 70.999^{+0.000070}_{-0.00010} $ & $ 67.9998^{+0.0093}_{-0.0082} $
       \\
       \\
       $ \alpha $ & $ 3.379^{+0.053}_{-0.039} $ & $ 3 \pm 0.00011 $ & $ 2.899^{+0.00014}_{-0.00012}  $ & $ 3.501^{+0.012}_{-0.0097}  $  
       \\
       \\
        $ \beta $ & $ 2.881 \pm 0.060 $ & $ 2.899994 ^{+0.000096}_{-0.000083} $ & $ 2.99999^{+0.000067}_{-0.00011}  $ & $ 1.9994 \pm 0.0091  $
      \\
      \\
       $ q_0 $ & -0.662831 & -0.695739 & -0.721428 & -0.324485
    \\
    \\
      $ z_{tr} $ & 0.723 &  0.731 &  0.740 &  0.449
     \\ 
     \\
\hline\hline  
\end{tabular}  
\end{center}
\end{table}

The jerk parameter $ (j) $ in terms of redshift $ z $ can be defined as \cite{Visser:2003vq, Liu:2023agr},
\begin{equation}
     j(z) = 1-2 \frac{1+z}{H(z)} H'(z) + \frac{(1+z)^2}{H(z)^2}(H'(z))^2 + \frac{(1+z)^2}{H(z)} H''(z) 
\end{equation}
Fig. \ref{jz} depicts the variation of the jerk parameter with $ z $. It is observed that the current values of the jerk parameter are approximated to be $ j_0 = 0.7783 $, $ 0.8388 $, $ 0.87143 $, and $ 0.5535 $ using OHD, Pantheon, OHD+Pantheon, and OHD+Pantheon+BAO datasets, respectively, that lie within limits over the cosmographic coefficients in the recent analysis \cite{Capozziello:2015rda}. In the future, it is approaching to 1.

The behavior of pressure due to matter is explained in Fig. (\ref{pz}) for all model parameters $ \alpha $, $ \beta $, and $ H_0 $. In this Figure, the pressure decreases as the value of redshift decreases and it becomes negative in the past and remains negative till late times. At present ($ z=0 $), the pressure is negative which indicates the presence of a mysterious force of dark energy that represents the expanding acceleration in the cosmos and according to that current analysis, the universe is in an accelerating phase. For different values of model parameters $ H_0 $, $ \alpha $ and $ \beta $, Fig. (\ref{rhoz}) explains the behavior of the matter density $ \rho $ over redshift $ z $.  There is a presence of a positive curve in the matter density which is approaching $ \infty $ as $ z \to \infty $. 

The equation of state (EoS) parameter is calculated as the ratio of the energy density ($ \rho $) and pressure due to matter ($ p $) that is $ \omega = \frac{p}{\rho} $. Fig. (\ref {wz}), explains the evolution of the EoS parameter over redshift $ z $ as it is approaching the $ \Lambda $CDM model which is the standard dark energy model in the future. It is observed that there is a transition from the perfect fluid model to a dust-free universe to the Quintessence model and then to a standard dark energy model that is $ \Lambda $CDM. Presently, it falls in the Quintessence region. The current values of the EoS parameter are approximated to be $ \omega_0 = -0.7752 $, $ -0.7971 $, $ -0.8142 $ and $ -0.5497 $ using OHD, Pantheon, OHD+Pantheon, and OHD+Pantheon+BAO datasets, respectively.

\subsection{Statefinder and $ Om(z) $ Diagnostic}
This section discusses the state finder diagnosis, a useful method to differentiate one dark energy model from another. There is an emergence of problems in discriminating the different dark energy models. To resolve this issue, Sahani et al. \cite{Sahni:2002fz} and Alam et al. \cite{Alam:2003sc} have developed a method that is termed a statefinder diagnostic. The state finder diagnostic involves the geometrical parameters pairs $ \{r, s\} $ and $ \{r, q\} $ where $ r $, $ s $ are the statefinder parameters respectively. So, the statefinder parameters depend on the expansion dynamics of the universe through high derivatives of the scale factor and are naturally one step beyond the Hubble parameter. The State finder pair $ \{r, s\} $ is defined as

\begin{figure}
    \begin{center}
       \subfloat[]{\label{sr} \includegraphics[scale=0.56]{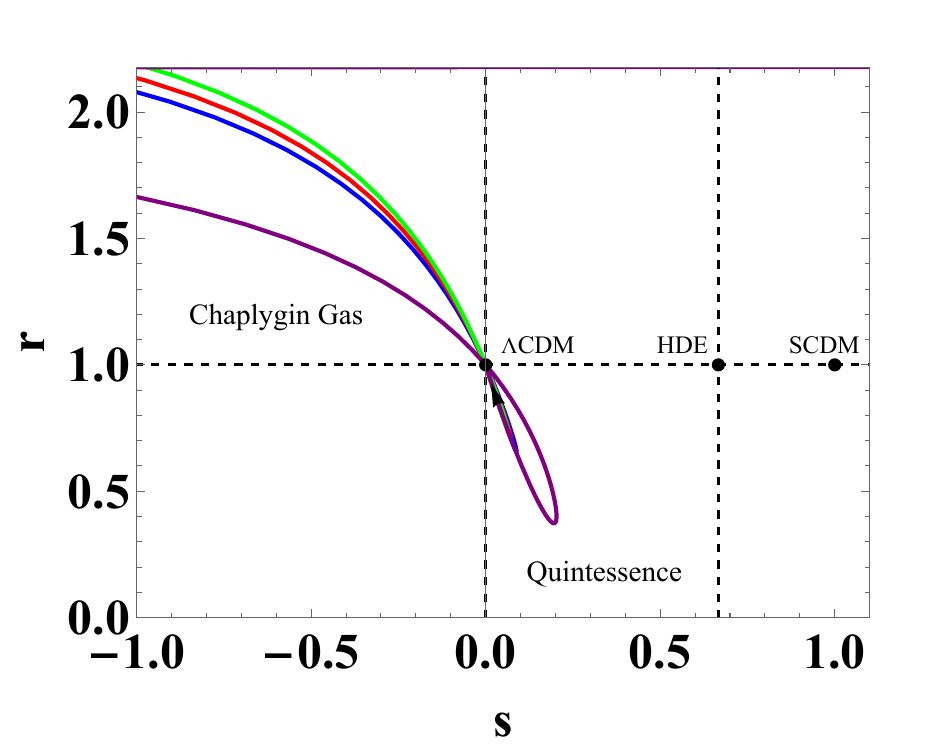}}\hfill
       \subfloat[]{\label{qr} \includegraphics[scale=0.54]{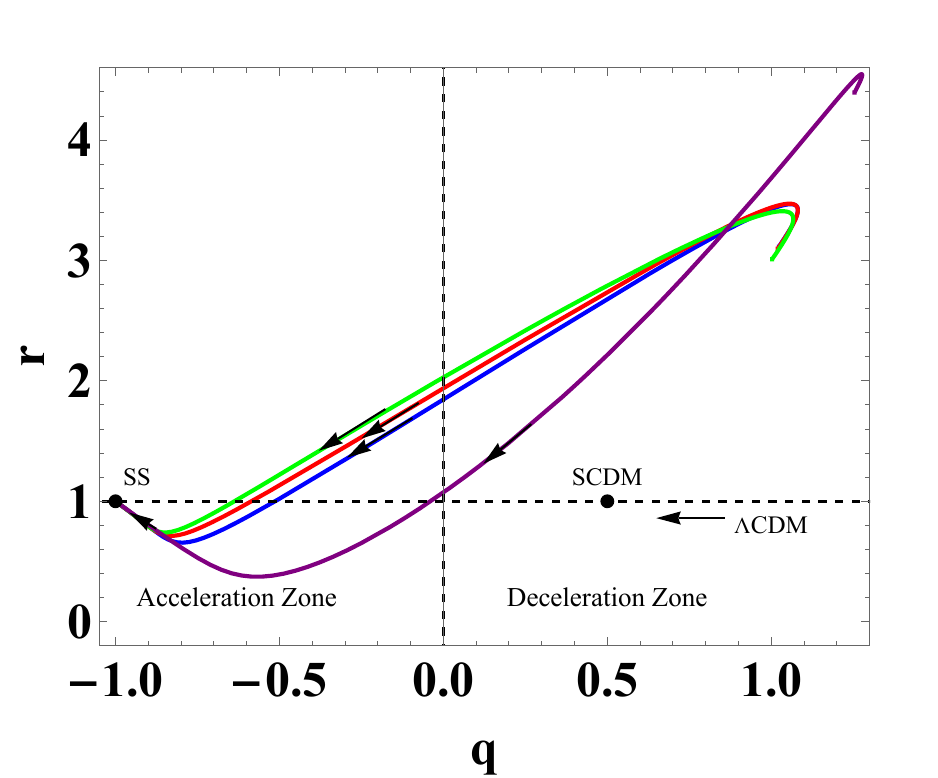}}
    \end{center}
\caption{{\it The statefinder plots in the planes of $ s-r $ and $ q-r $ for the various observational datasets.}}
\label{sq}
\end{figure}

\begin{equation}{\label{24}}
    r=\frac{\dddot{a}}{aH^3}, \hskip 0.2in s=\frac{r-1}{3(q-\frac{1}{2})},
\end{equation}
where $ q \neq \frac{1}{2} $ is decelerating parameter. The method is considered effective as the fixed point $ (0, 1) $ and  $ (1, 1) $ in the plot of $ \{s, r\} $ represents the $ \Lambda $CDM  model and the $ SCDM $ model respectively. And  $ (-1, 1) $ represents the steady-state model ($ SS $) in $ \{q, r\} $ plot. This analysis helps in distinguishing various dark energy models such as the Chaplygin gas Model, Quintessence Model, and Brane world DE model. In Fig. (\ref{qr}), it is observed that both models are approaching the Steady-state model. In Fig. (\ref{sr}), it is observed that all models approach the standard $ \Lambda $CDM. It initially lies in Chaplygin Gas and then lies in the Quintessence model space. The $ Om(z) $ is a diagnostic of dark energy proposed in \cite{Sahni:2008xx} which is constructed from the Hubble parameter $ H(z) $ and defined as,

\begin{figure}[tbph]
    \begin{center}
    \includegraphics[scale=0.55]{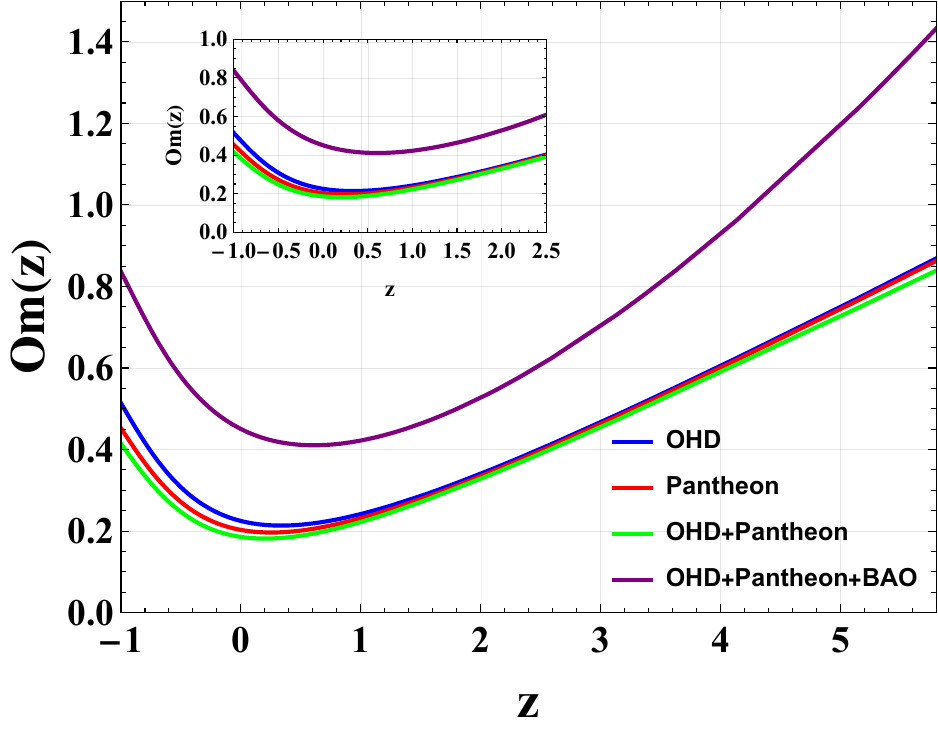}
    \end{center}
\caption{{\it The $ Om (z) $ diagnostics plots for the various observational datasets.}}
\label{Om(z)}
\end{figure}

\begin{equation}
    Om(z) = \frac{ \bigg( \frac{H(z)}{H_0} \bigg)^2 -1}{(1+z)^3-1}.
\end{equation}

The negative, null, and positive slope of the $ Om $ indicates the quintessence-like behavior, $ \Lambda $CDM, and phantom-like behavior respectively. The evolution of Om(z) is shown in Fig. \ref{Om(z)}. It is noticed that the Om(z) has a negative slope for $ z < 0.3 $ implying that the dark energy of our model displays a quintessence-like behavior, which is consistent with the evolution of the dark energy EoS, $ \omega $ in the corresponding redshift range $ z < 0.3 $ (see Fig. \ref{wz}) \cite{Acero:2024jqe}. Therefore, the $ f(R, G) $ model demonstrates the quintessence-like behavior.

\section{Energy Condition}{\label{sec-6}}
Modifications to GR, such as $ f(R, G) $ gravity, can be used in different contexts to explain many cosmological events. But it needs testing criteria to check the viability of these models, one of them is energy conditions. This section discusses the energy conditions that play an important role in understanding the geodesics of the universe. Visser \cite{Visser:1997qk} has discussed four types of energy conditions in GR as given in Table \ref{tab:Energy Conditions}:

\begin{figure}
    \begin{center}
        \subfloat[]{\label{NEC} \includegraphics[scale=0.55]{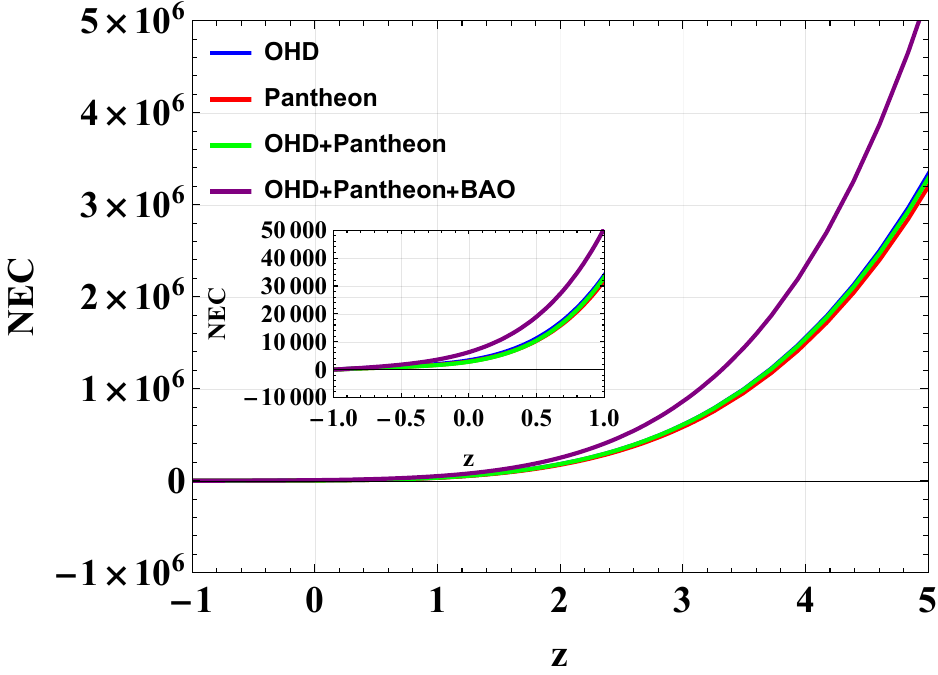}}\hfill
        \subfloat[]{\label{DEC} \includegraphics[scale=0.55]{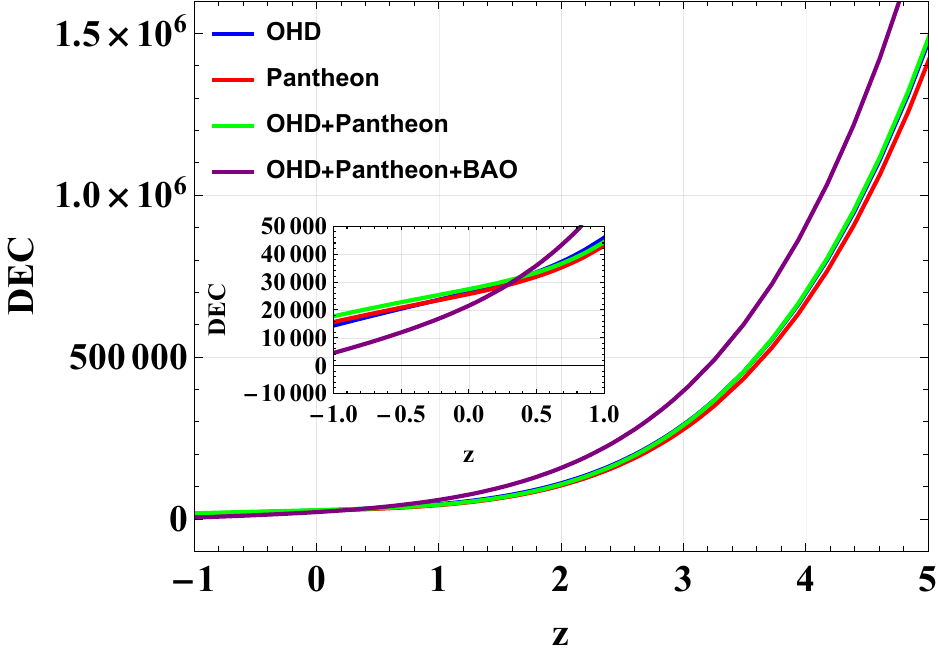}}\par
        \subfloat[]{\label{SEC} \includegraphics[scale=0.55]{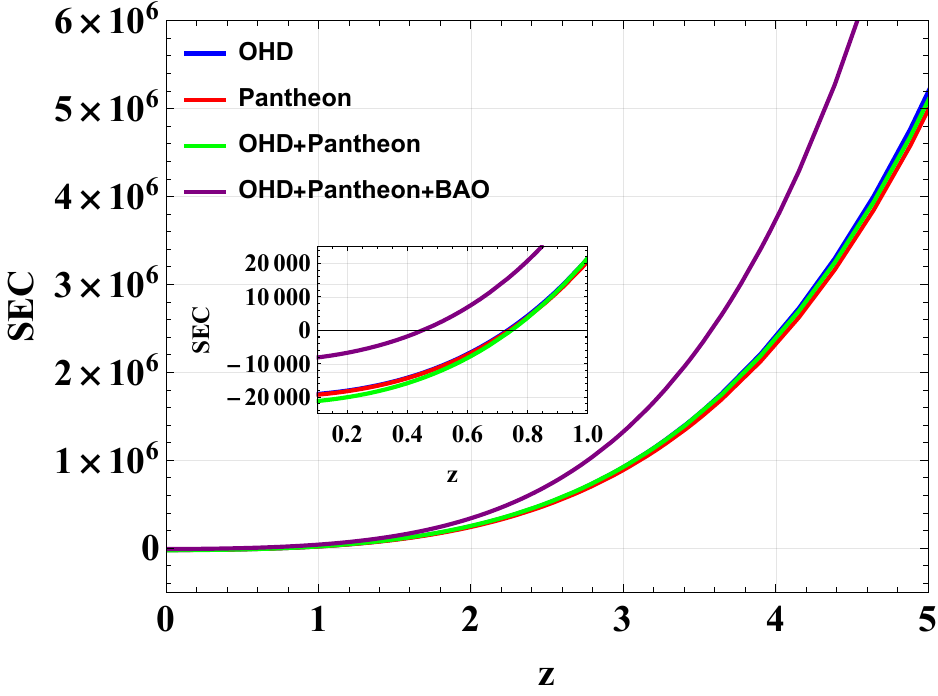}}
    \end{center}
   \caption{{\it The variations of the Energy Conditions.}}
    \label{EC}
\end{figure}

In fig. \ref{EC}, It is observed that the model satisfies the energy conditions NEC and DEC but does not satisfy SEC in the later times which indicates the existence of dark energy at later times. 

\begin{table}[htbp]
\caption{ \textbf{Energy Conditions}}
\centering
\begin{tabular}{l c c c c r}
\hline\hline
\\
Energy Condition & \qquad Physical form & \qquad Geometric form & \qquad Perfect Fluid form
\\
 \\
\hline
\\
 ~~ $ NEC $ ~~ & ~~ $ T_{ij}k^ik^j\geq 0 $ & ~~$ R_{ij}k^ik^j\geq 0 $  &  ~~ $ \rho+p \geq 0 $
\\
\\ 
~~ $ WEC $ ~~ & ~~ $ T_{ij}t^it^j\geq 0 $  & ~~$  G_{ij}t^it^j\geq 0 $  & ~~$  \rho \geq 0, \rho+p \geq 0 $  
 \\
\\
~~ $ SEC $ ~~ & ~~  $ (T_{ij}-\frac{T}{n-2}g_{ij})t^it^j\geq 0  $ & ~~ $ R_{ij}t^it^j\geq 0 $  & ~~  $ \rho+p \geq 0, (n-3)\rho+(n-1)p \geq 0 $
\\
\\
~~ $ DEC $~~  & ~~ $ T_{ij}t^i\xi^j\geq 0 $  & ~~ $ G_{ij}t^i\xi^j\geq 0 $ & ~~   $\rho \geq |p| $  
\\
 \\
\hline\hline
\end{tabular}
\label{tab:Energy Conditions}
\end{table}

\section{Thermodynamic Analysis}{\label{sec-7}}

This section discusses the generalized second law (GSL) of thermodynamics validity for the model \cite{Singh:2023bjx}. The second law of thermodynamics states that the Universe's total entropy increases with respect to time. The total entropy ($ S $) is equal to the sum of the entropy contributed due to matter ($ S_m $) and apparent horizon ($ S_h $), i.e. $ S = S_m + S_h $, where $ S_h = \frac{\kappa_{B} A}{4 {l^2}_{Pl}} $,  $ l_{Pl} $  and $ \kappa_{B} $ denotes the Planck's length and the Boltzman constant respectively \cite{Cai:2008gw}. $ A $ denotes the area of the horizon with radius $ r_h = \frac{1}{\sqrt{(H^2 + k a^{-2})}} $
As we are considering spatially flat model ($ k=0 $), therefore $ r_h = \frac{1}{H} $  and then horizon entropy will become, 

\begin{equation}
     S_h = \frac{\kappa_B \pi}{ {l^2}_{Pl} H^2},
\end{equation}

\begin{equation}
     T d S_m = d(\rho V) + p dV = V d(\rho) + ( \rho + p) dV,
\end{equation}
where $ V = \frac{4 \pi }{3} {r^3}_h $ is the spatial volume enclosed by the horizon and $ T $ is the temperature of the fluid, and $ T $ is assumed to be equal to the temperature at the horizon $ T_h $, where $ T_h = \frac{1}{2 \pi r_h} $ \cite{Akbar:2006kj}.

\begin{figure}[tbph]
    \begin{center}
        \subfloat[]{\label{temp} \includegraphics[scale=0.53]{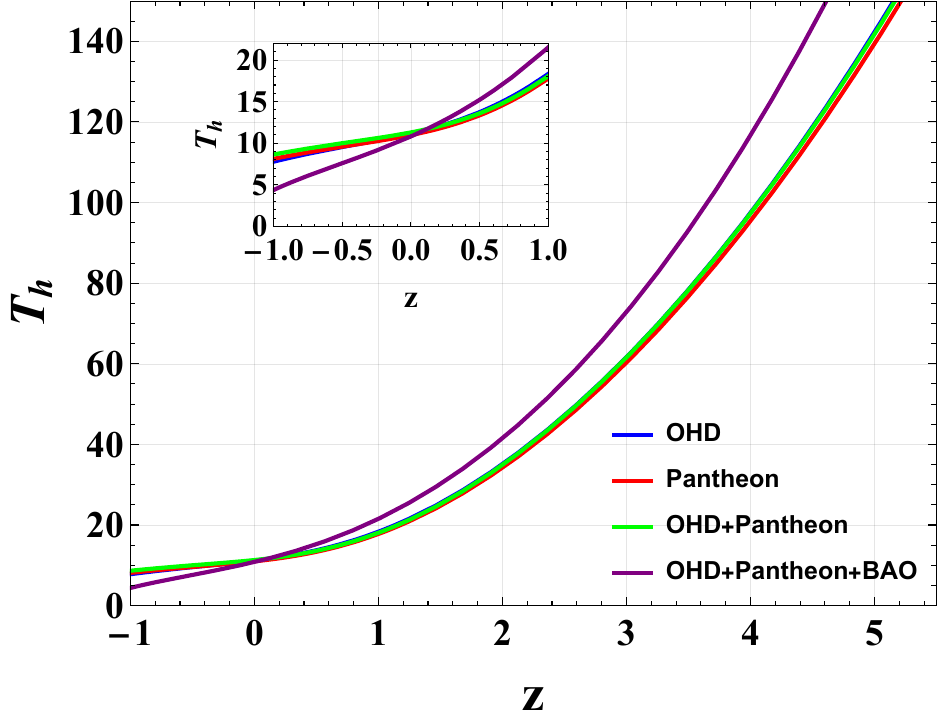}}\hfill
        \subfloat[]{\label{Total S} \includegraphics[scale=0.54]{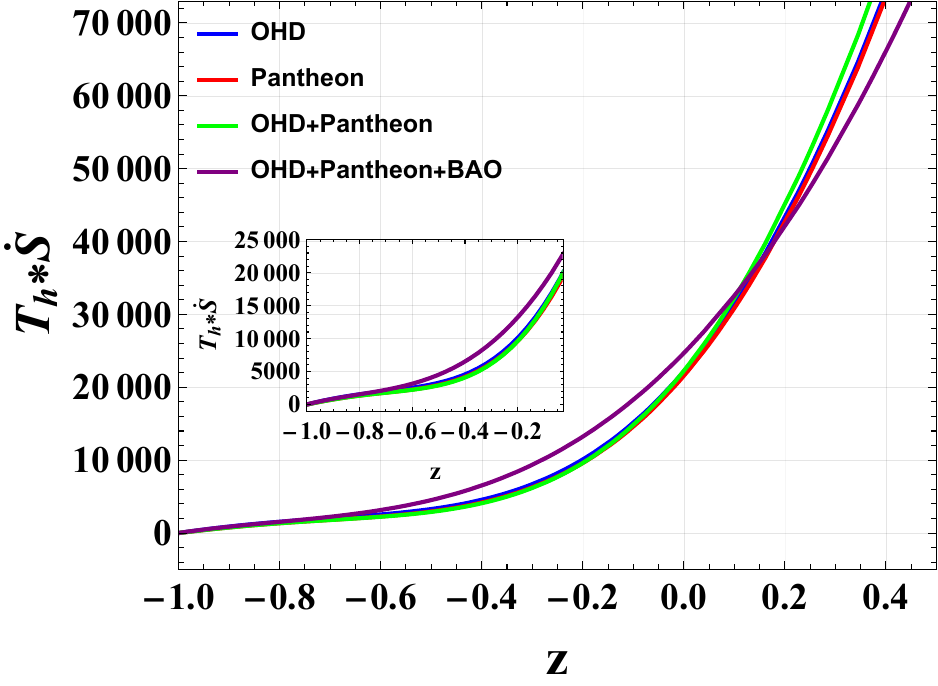}}
    \end{center}
\caption{{\it The evolution of the Hawking temperature $ T_h $ and the total entropy $ T_h \dot{S} $  with respect to the redshift $ z $ for the various observational datasets.}}
\label{TE}
\end{figure}

In Fig. \ref{temp}, it is observed that the Hawking temperature $ T_h $ is positive for all values of redshift and it increases with an increase in redshift. Fig. \ref{Total S} exhibits that the total entropy $  T_h \dot{S} $ is positive for the entire redshift range which shows that this model satisfies the second law of thermodynamics.

\section{Conclusion}{\label{sec-8}}
We have studied the behavior of the universe in a modified form of gravity $ f(R, G) $, where $ G $ is the Gauss-Bonnet invariant consisting of the higher order curvature term in $ R $. We adopted a simple power law function of $ f(R, G) $ as $ R + f(G) $, where $ f(G) = G^{1-m} $, $ m $ is an arbitrary constant taken to be $ m > 1 $.  The gravitational field equations have been solved by parametrizing the Hubble parameter $ H $ and obtaining appropriate solutions for various physical parameters. To obtain the optimal values of the model parameters $ \alpha $, $ \beta $, and current Hubble parameter $ H_0 $, we use the Markov Chain Monte Carlo (MCMC) method by applying the emcee codes in Python using various observational datasets discussed in the Figs. \ref{Hz}. The error bar plots of the Hubble parameter $ H(z) $, distance modulus $ \mu(z) $, and the apparent magnitudes $ m_b(z) $ exhibit the degree of consistency between the presented model and the $ \Lambda $CDM (see Figs. \ref{err}). 

The current values of the cosmological parameters  $ H $  and $ q $ along with model parameters for different observational datasets are evaluated in Table \ref{tab: parameters value}. We assessed the behavior of the deceleration parameter for the evaluated best-fit values in Fig. \ref{qz}. There is a transition from the positive values to negative values, and the transition nodes vary in the range of redshift $ 0.43 < z_{tr} < 0.76 $. At present, our model is in the accelerating phase of the universe. The jerk parameter is positive for all the redshift ranges and converges to $1$ at later times for all the constrained model parameters shown in Fig. \ref{jz}. The evolution of the energy density is always positive and decreases as the cosmic redshift decreases (see Fig. \ref{rhoz}). 

The isotropic pressure of the model is positive for the redshift ranges $ z>2.341 $, shows the dust-filled Universe at the redshift $ z\asymp2.341 $, and demonstrates the negative behavior for all the constraints values of the model parameters in later times. The model indicates the existence of dark energy in the universe that leads to the accelerating expansion of the universe in later times (see Fig. \ref{pz}). The EoS parameter transits from positive to negative values which approaches $ -1 $ shown in the Fig. \ref{wz}. Hence the the model behaves like the quintessence dark energy model and converges to the the standard dark energy model $ \Lambda $CDM at later times.

Fig. \ref{sq} depicts the behaviors of the Statefinder Diagnostic pairs $ \{s, r\} $ and $ \{q, r\} $. The model starts with the Chaplygin region, enters into the quintessence region, and approaches the $ \Lambda $CDM in the parametric space $ \{s, r\} $. The model transits from the deceleration zone to the acceleration zone, deviates from SCDM, crosses the Quintom line, and approaches the de Sitter state in the parametric space $ \{q, r\} $. The evolution of Om(z) is shown in Fig. \ref{Om(z)}. It is noticed that the Om(z) has a negative slope for $ z < 0.3 $ implying that the dark energy of our model displays a quintessence-like behavior, which is consistent with the evolution of the dark energy EoS, $ \omega $ in the corresponding redshift range $ z < 0.3 $ (see Fig. \ref{wz}). Therefore, the $ f(R, G) $ model demonstrates the quintessence-like behavior.

In fig. \ref{EC}, It is observed that the model satisfies the energy conditions NEC and DEC but does not satisfy SEC in the later times which indicates the existence of dark energy at later times. Additionally,  It is observed that the Hawking temperature $ T_h $ and the total entropy $  T_h \dot{S} $ are positive for the entire redshift range and decrease in later times. Thus, our model satisfies the second law of thermodynamics (see Fig. \ref{TE}). Finally, it is concluded that the $ f(R, G) $ gravity model within the framework of FLRW metric is consistent with the recent observational datasets and behaves as a quintessence-like accelerating expansion model of the universe.



\begin{thebibliography}{9999}

\bibitem{SupernovaSearchTeam:1998fmf}
A.~G.~Riess \textit{et al.} [Supernova Search Team],
Astron. J. \textbf{116}, 1009-1038 (1998).

\bibitem{SupernovaCosmologyProject:1998vns}
S.~Perlmutter \textit{et al.} [Supernova Cosmology Project],
Astrophys. J. \textbf{517}, 565-586 (1999).


\bibitem{Perlmutter:1999jt}
S.~Perlmutter, M.~S.~Turner and M.~J.~White,
Phys. Rev. Lett. \textbf{83}, 670-673 (1999).


\bibitem{WMAP:2003elm}
D.~N.~Spergel \textit{et al.} [WMAP],
Astrophys. J. Suppl. \textbf{148}, 175-194 (2003).

%
\bibitem{WMAP:2006bqn}
D.~N.~Spergel \textit{et al.} [WMAP],
Astrophys. J. Suppl. \textbf{170}, 377 (2007).

\bibitem{WMAP:2010qai}
E.~Komatsu \textit{et al.} [WMAP],
Astrophys. J. Suppl. \textbf{192}, 18 (2011).

\bibitem{Planck:2018vyg}
N.~Aghanim \textit{et al.} [Planck],
Astron. Astrophys. \textbf{641}, A6 (2020).

\bibitem{Sahni:1999gb}
V.~Sahni and A.~A.~Starobinsky,
Int. J. Mod. Phys. D \textbf{9}, 373-444 (2000).

\bibitem{Chevallier:2000qy}
M.~Chevallier and D.~Polarski,
Int. J. Mod. Phys. D \textbf{10}, 213-224 (2001).

\bibitem{Carroll:2000fy}
S.~M.~Carroll,
Living Rev. Rel. \textbf{4}, 1 (2001).

\bibitem{Copeland:2006wr}
E.~J.~Copeland, M.~Sami and S.~Tsujikawa,
Int. J. Mod. Phys. D \textbf{15}, 1753-1936 (2006).

\bibitem{Frieman:2008sn}
J.~Frieman, M.~Turner and D.~Huterer,
Ann. Rev. Astron. Astrophys. \textbf{46}, 385-432 (2008).

\bibitem{amendola2010dark}   
Amendola, Luca and Tsujikawa, Shinji
Cambridge University Press (2010).


\bibitem{Nojiri:2006ri}
S.~Nojiri and S.~D.~Odintsov,
eConf \textbf{C0602061}, 06 (2006).

\bibitem{Capozziello:2007ec}
S.~Capozziello and M.~Francaviglia,
Gen. Rel. Grav. \textbf{40}, 357-420 (2008).

\bibitem{Lobo:2008sg}
F.~S.~N.~Lobo,
[arXiv:0807.1640 [gr-qc]].

\bibitem{deHaro:2023lbq}
J.~de Haro, S.~Nojiri, S.~D.~Odintsov, V.~K.~Oikonomou and S.~Pan,
Phys. Rept. \textbf{1034}, 1-114 (2023).

\bibitem{Singh:2018xjv}
J.~K.~Singh, K.~Bamba, R.~Nagpal and S.~K.~J.~Pacif,
Phys. Rev. D \textbf{97}, no.12, 123536 (2018).


\bibitem{Aviles:2014rma}
A.~Aviles, A.~Bravetti, S.~Capozziello and O.~Luongo,
Phys. Rev. D \textbf{90}, no.4, 043531 (2014).

\bibitem{delaCruz-Dombriz:2016bqh}
\'A.~de la Cruz-Dombriz, P.~K.~S.~Dunsby, O.~Luongo and L.~Reverberi,
JCAP \textbf{12}, 042 (2016).

\bibitem{Capozziello:2019cav}
S.~Capozziello, R.~D'Agostino and O.~Luongo,
Int. J. Mod. Phys. D \textbf{28}, no.10, 1930016 (2019).


\bibitem{Shaily:2024nmy}
Shaily, A.~Singh, J.~K.~Singh and S.~Ray,
Astron. Comput. \textbf{49} (2024), 100876.

\bibitem{Balhara:2023mgj}
H.~Balhara, J.~K.~Singh and E.~N.~Saridakis,
[arXiv:2312.17277 [gr-qc]].


\bibitem{Shaily:2024xho}
Shaily, A.~Singh, J.~K.~Singh, S.~Hussain and R.~Myrzakulov,
[arXiv:2402.08709 [gr-qc]].


\bibitem{Pawar:2024juv}
D.~D.~Pawar, D.~K.~Raut, A.~P.~Nirwal, Shaily and J.~K.~Singh,
Astron. Comput. \textbf{48} (2024), 100848.

\bibitem{Shabani:2016dhj}
H.~Shabani and A.~H.~Ziaie,
Eur. Phys. J. C \textbf{77} (2017) no.1, 31.

\bibitem{Singh:2022nfm}
J.~K.~Singh, Shaily, S.~Ram, J.~R.~L.~Santos and J.~A.~S.~Fortunato,
Int. J. Mod. Phys. D \textbf{32}, no.07, 2350040 (2023).


\bibitem{Singh:2023bjx}
J.~K.~Singh, Shaily, A.~Pradhan and A.~Beesham,
[arXiv:2304.09917 [gr-qc]].

\bibitem{Singh:2024gtz}
J.~K.~Singh, Shaily, A.~Singh, H.~Balhara and J.~R.~L.~Santos,
Annals Phys. \textbf{469} (2024), 169781.

\bibitem{Singh:2023yye}
J.~K.~Singh and R.~Nagpal,
Indian J. Phys. \textbf{98} (2024) no.7, 2609-2622.





\bibitem{Nojiri:2003ft}
S.~Nojiri and S.~D.~Odintsov,
Phys. Rev. D \textbf{68}, 123512 (2003).


\bibitem{Carroll:2003wy}
S.~M.~Carroll, V.~Duvvuri, M.~Trodden and M.~S.~Turner,
Phys. Rev. D \textbf{70}, 043528 (2004).

\bibitem{Chiba:2006jp}
T.~Chiba, T.~L.~Smith and A.~L.~Erickcek,
Phys. Rev. D \textbf{75}, 124014 (2007).

\bibitem{Sotiriou:2008rp}
T.~P.~Sotiriou and V.~Faraoni,
Rev. Mod. Phys. \textbf{82}, 451-497 (2010).

\bibitem{Starobinsky:2007hu}
A.~A.~Starobinsky,
JETP Lett. \textbf{86}, 157-163 (2007).

\bibitem{Bertolami:2007gv}
O.~Bertolami, C.~G.~Boehmer, T.~Harko and F.~S.~N.~Lobo,
Phys. Rev. D \textbf{75}, 104016 (2007).

\bibitem{Capozziello:2008qc}
S.~Capozziello, V.~F.~Cardone and V.~Salzano,
Phys. Rev. D \textbf{78}, 063504 (2008).


\bibitem{Nojiri:2010wj}
S.~Nojiri and S.~D.~Odintsov,
Phys. Rept. \textbf{505}, 59-144 (2011).

\bibitem{Goswami:2022vfq}
G.~K.~Goswami, R.~Rani, H.~Balhara and J.~K.~Singh,
Indian J. Phys. \textbf{97}, no.12, 3707-3714 (2023).

\bibitem{Harko:2011kv}
T.~Harko, F.~S.~N.~Lobo, S.~Nojiri and S.~D.~Odintsov,
Phys. Rev. D \textbf{84}, 024020 (2011).

\bibitem{Capozziello:2014bqa}
S.~Capozziello, F.~S.~N.~Lobo and J.~P.~Mimoso,
Phys. Rev. D \textbf{91}, no.12, 124019 (2015).

\bibitem{Singh:2022jue}
J.~K.~Singh, H.~Balhara, K.~Bamba and J.~Jena,
JHEP \textbf{03}, 191 (2023)
[erratum: JHEP \textbf{04}, 049 (2023)].


\bibitem{Pradhan:2023}
A.~Pradhan, G.~Goswami, R.~Rani and A.~Beesham,
[arXiv:2210.15433 [gr-qc]].
Astronomy and Computing \textbf{44}, 100737(2023).

\bibitem{Singh:2023gxd}
J.~K.~Singh, Shaily, A.~Singh, A.~Beesham and H.~Shabani,
Annals Phys. \textbf{455}, 169382 (2023).


\bibitem{Singh:2024ckh}
J.~K.~Singh, Shaily, H.~Balhara, S.~G.~Ghosh and S.~D.~Maharaj,
Phys. Dark Univ. \textbf{45} (2024), 101513.

\bibitem{Singh:2024kez}
J.~K.~Singh, H.~Balhara, Shaily and P.~Singh,
Astron. Comput. \textbf{46}, 100795 (2024).



\bibitem{Ferraro:2006jd}
R.~Ferraro and F.~Fiorini,
Phys. Rev. D \textbf{75}, 084031 (2007).

\bibitem{Cai:2015emx}
Y.~F.~Cai, S.~Capozziello, M.~De Laurentis and E.~N.~Saridakis,
Rept. Prog. Phys. \textbf{79}, no.10, 106901 (2016).


\bibitem{Paliathanasis:2016vsw}
A.~Paliathanasis, J.~D.~Barrow and P.~G.~L.~Leach,
Phys. Rev. D \textbf{94}, no.2, 023525 (2016).

\bibitem{Duchaniya:2023aeu}
L.~K.~Duchaniya, K.~Gandhi and B.~Mishra,
Phys. Dark Univ. \textbf{44}, 101461 (2024).



\bibitem{Townsend:1979js}
P.~K.~Townsend and P.~van Nieuwenhuizen,
Phys. Rev. D \textbf{19}, 3592 (1979).

\bibitem{Birrell_1982}
N.~ D.~ Birrell and P.~ C.~ W.~ Davies, 
Cambridge University Press, Cambridge, UK, (1982); 

\bibitem{Barth:1983hb}
N.~H.~Barth and S.~M.~Christensen,
Phys. Rev. D \textbf{28}, 1876 (1983).


\bibitem{DeFelice:2008wz}
A.~De Felice and S.~Tsujikawa,
Phys. Lett. B \textbf{675}, 1-8 (2009).


\bibitem{MontelongoGarcia:2010ip}
N.~Montelongo Garcia, F.~S.~N.~Lobo, J.~P.~Mimoso and T.~Harko,
J. Phys. Conf. Ser. \textbf{314}, 012056 (2011).


\bibitem{Sharif:2016drh}
M.~Sharif and H.~I.~Fatima,
Gen. Rel. Grav. \textbf{49}, 1, 1 (2017).


\bibitem{Boehmer:2021aji}
C.~G.~Boehmer and E.~Jensko,
Phys. Rev. D \textbf{104}, no.2, 024010 (2021).


\bibitem{Mandal:2020lyq}
S.~Mandal, P.~K.~Sahoo and J.~R.~L.~Santos,
Phys. Rev. D \textbf{102}, no.2, 024057 (2020).


\bibitem{Mandal:2020buf}
S.~Mandal, D.~Wang and P.~K.~Sahoo,
Phys. Rev. D \textbf{102}, 124029 (2020).

\bibitem{Frusciante:2021sio}
N.~Frusciante,
Phys. Rev. D \textbf{103}, no.4, 044021 (2021).

\bibitem{Heisenberg:2023lru}
L.~Heisenberg,
Phys. Rept. \textbf{1066}, 1-78 (2024).

\bibitem{Goswami:2023knh}
G.~K.~Goswami, R.~Rani, J.~K.~Singh and A.~Pradhan,
JHEAp \textbf{43} (2024), 105-113.


\bibitem{Nojiri:2005jg}
S.~Nojiri and S.~D.~Odintsov,
Phys. Lett. B \textbf{631}, 1-6 (2005).

\bibitem{Nojiri:2005am}
S.~Nojiri, S.~D.~Odintsov and O.~G.~Gorbunova,
J. Phys. A \textbf{39}, 6627-6634 (2006).


\bibitem{Li:2007jm}
B.~Li, J.~D.~Barrow and D.~F.~Mota,
Phys. Rev. D \textbf{76}, 044027 (2007).


%
\bibitem{Achucarro:1986uwr}
A.~Achucarro and P.~K.~Townsend,
Phys. Lett. B \textbf{180}, 89 (1986).

\bibitem{Gomez:2011zzd}
F.~Gomez, P.~Minning and P.~Salgado,
Phys. Rev. D \textbf{84}, 063506 (2011).


\bibitem{Raushan:2023pdv}
R.~Raushan and A.~Singh,
Phys. Dark Univ. \textbf{39}, 101152 (2023).

\bibitem{Lovelock:1971yv}
D.~Lovelock,
J. Math. Phys. \textbf{12}, 498-501 (1971).

\bibitem{Bajardi:2021hya}
F.~Bajardi, D.~Vernieri and S.~Capozziello,
JCAP \textbf{11}, no.11, 057 (2021).



\bibitem{Leigh:1989jq}
R.~G.~Leigh,
Mod. Phys. Lett. A \textbf{4}, 2767 (1989).

\bibitem{Tseytlin:1997csa}
A.~A.~Tseytlin,
Nucl. Phys. B \textbf{501}, 41-52 (1997).



\bibitem{Atazadeh:2013cz}
K.~Atazadeh and F.~Darabi,
Gen. Rel. Grav. \textbf{46}, 1664 (2014).

\bibitem{DeLaurentis:2015fea}
M.~De Laurentis, M.~Paolella and S.~Capozziello,
Phys. Rev. D \textbf{91}, no.8, 083531 (2015).


\bibitem{Odintsov:2018nch}
S.~D.~Odintsov, V.~K.~Oikonomou and S.~Banerjee,
Nucl. Phys. B \textbf{938}, 935-956 (2019).


\bibitem{Bhatti:2020cjz}
M.~Z.~Bhatti, Z.~Yousaf and A.~Rehman,
Phys. Dark Univ. \textbf{29}, 100561 (2020).


\bibitem{Lohakare:2021yuo}
S.~V.~Lohakare, S.~K.~Tripathy and B.~Mishra,
Phys. Scripta \textbf{96}, no.12, 125039 (2021).

\bibitem{Shekh:2021dsl}
S.~H.~Shekh,
New Astron. \textbf{83}, 101464 (2021).

\bibitem{Chanda:2023lfk}
A.~Chanda, A.~Chanda, A.~Halder, A.~Haldar, A.~S.~Majumdar, A.~S.~Majumdar, B.~C.~Paul and B.~C.~Paul,
Eur. Phys. J. C \textbf{83}, no.1, 23 (2023).


\bibitem{Shaily:2024rjq}
Shaily, J.~K.~Singh and A.~Singh,
Fortsch. Phys. \textbf{72}, no.6, 2300244 (2024).

\bibitem{Das:2023bff}
K.~P.~Das, U.~Debnath, A.~Ashraf and M.~Khurana,
Phys. Dark Univ. \textbf{43}, 101398 (2024).


\bibitem{Bamba:2010wfw}
K.~Bamba, S.~D.~Odintsov, L.~Sebastiani and S.~Zerbini,
Eur. Phys. J. C \textbf{67}, 295-310 (2010).


\bibitem{DeFelice:2011ka}
A.~De Felice, T.~Suyama and T.~Tanaka,
Phys. Rev. D \textbf{83}, 104035 (2011).

\bibitem{Odintsov:2015uca}
S.~D.~Odintsov, V.~K.~Oikonomou and E.~N.~Saridakis,
Annals Phys. \textbf{363}, 141-163 (2015).


\bibitem{Wu:2015maa}
B.~Wu and B.~Q.~Ma,
Phys. Rev. D \textbf{92}, no.4, 044012 (2015).


\bibitem{Lansberg:2016deg}
J.~P.~Lansberg and H.~S.~Shao,
Eur. Phys. J. C \textbf{77}, no.1, 1 (2017).


\bibitem{Calza:2019egu}
M.~Calz\`a, A.~Casalino, O.~Luongo and L.~Sebastiani,
Eur. Phys. J. Plus \textbf{135}, no.1, 1 (2020).


\bibitem{Elizalde:2020zcb}
E.~Elizalde, S.~D.~Odintsov, V.~K.~Oikonomou and T.~Paul,
Nucl. Phys. B \textbf{954}, 114984 (2020).

\bibitem{Mustafa:2020jln}
G.~Mustafa, M.~F.~Shamir and X.~Tie-Cheng,
Phys. Rev. D \textbf{101}, no.10, 104013 (2020).

\bibitem{Nada:2020jay}
A.~Nada and A.~Ramos,
Eur. Phys. J. C \textbf{81}, no.1, 1 (2021).

\bibitem{Naz:2023pfl}
T.~Naz, A.~Malik, M.~K.~Asif and I.~Fayyaz,
Phys. Dark Univ. \textbf{42}, 101301 (2023).




\bibitem{Singh:2022gln}
J.~K.~Singh, Shaily and K.~Bamba,
Chin. J. Phys. \textbf{84}, 371-380 (2023).


\bibitem{Myrzakulov:2023sir}
N.~Myrzakulov, M.~Koussour and A.~Mussatayeva,
Chin. J. Phys. \textbf{85}, 345-358 (2023).


\bibitem{Foreman-Mackey:2012any}
D.~Foreman-Mackey, D.~W.~Hogg, D.~Lang and J.~Goodman,
Publ. Astron. Soc. Pac. \textbf{125}, 306-312 (2013).


\bibitem{Shaily:2022enj}
Shaily, J.~K.~Singh, J.~R.~L.~Santos and M.~Zeyauddin,
Int. J. Mod. Phys. D \textbf{33} (2024), 2450024.

\bibitem{Padmanabhan:2012hf}
N.~Padmanabhan, X.~Xu, D.~J.~Eisenstein, R.~Scalzo, A.~J.~Cuesta, K.~T.~Mehta and E.~Kazin,
Mon. Not. Roy. Astron. Soc. \textbf{427}, no.3, 2132-2145 (2012).

\bibitem{Beutler:2011hx}
F.~Beutler, C.~Blake, M.~Colless, D.~H.~Jones, L.~Staveley-Smith, L.~Campbell, Q.~Parker, W.~Saunders and F.~Watson,
Mon. Not. Roy. Astron. Soc. \textbf{416}, 3017-3032 (2011).


\bibitem{BOSS:2013rlg}
L.~Anderson \textit{et al.} [BOSS],
Mon. Not. Roy. Astron. Soc. \textbf{441}, no.1, 24-62 (2014).

%
\bibitem{Blake:2012pj}
C.~Blake, S.~Brough, M.~Colless, C.~Contreras, W.~Couch, S.~Croom, D.~Croton, T.~Davis, M.~J.~Drinkwater and K.~Forster, \textit{et al.}
Mon. Not. Roy. Astron. Soc. \textbf{425}, 405-414 (2012).


\bibitem{Blake:2011en}
C.~Blake, E.~Kazin, F.~Beutler, T.~Davis, D.~Parkinson, S.~Brough, M.~Colless, C.~Contreras, W.~Couch and S.~Croom, \textit{et al.}
Mon. Not. Roy. Astron. Soc. \textbf{418}, 1707-1724 (2011).


\bibitem{SDSS:2009ocz}
W.~J.~Percival \textit{et al.} [SDSS],
Mon. Not. Roy. Astron. Soc. \textbf{401}, 2148-2168 (2010).

\bibitem{WMAP:2012nax}
G.~Hinshaw \textit{et al.} [WMAP],
Astrophys. J. Suppl. \textbf{208}, 19 (2013).


\bibitem{SDSS:2005xqv}
D.~J.~Eisenstein \textit{et al.} [SDSS],
Astrophys. J. \textbf{633}, 560-574 (2005).

\bibitem{giostri2012cosmic}
R.~Giostri, M.~V.~dSantos, I.~Waga, R.~R.~R.~Reis, M.~O.~Calvao, B.~L.~Lago,
J. Cosmol. Astropart. Phys, \textbf{1203}, 027 (2012).


\bibitem{Visser:2003vq}
M.~Visser,
Class. Quant. Grav. \textbf{21}, 2603-2616 (2004).


\bibitem{Liu:2023agr}
J.~Liu, L.~Qiao, B.~Chang and L.~Xu,
Eur. Phys. J. C \textbf{83}, no.5, 374 (2023).


\bibitem{Capozziello:2014zda}
S.~Capozziello, O.~Farooq, O.~Luongo and B.~Ratra,
Phys. Rev. D \textbf{90}, no.4, 044016 (2014).


\bibitem{Capozziello:2015rda}
S.~Capozziello, O.~Luongo and E.~N.~Saridakis,
Phys. Rev. D \textbf{91}, no.12, 124037 (2015).


\bibitem{Sahni:2002fz}
V.~Sahni, T.~D.~Saini, A.~A.~Starobinsky and U.~Alam,
JETP Lett. \textbf{77}, 201-206 (2003).

\bibitem{Alam:2003sc}
U.~Alam, V.~Sahni, T.~D.~Saini and A.~A.~Starobinsky,
Mon. Not. Roy. Astron. Soc. \textbf{344}, 1057 (2003).

\bibitem{Sahni:2008xx}
V.~Sahni, A.~Shafieloo and A.~A.~Starobinsky,
Phys. Rev. D \textbf{78}, 103502 (2008).

\bibitem{Acero:2024jqe}
M.~A.~Acero and A.~Oliveros,
[arXiv:2406.00207 [astro-ph.CO]].

\bibitem{Visser:1997qk}
M.~Visser,
Science \textbf{276}, 88-90 (1997).

\bibitem{Cai:2008gw}
R.~G.~Cai, L.~M.~Cao and Y.~P.~Hu,
Class. Quant. Grav. \textbf{26}, 155018 (2009).

\bibitem{Akbar:2006kj}
M.~Akbar and R.~G.~Cai,
Phys. Rev. D \textbf{75}, 084003 (2007).


\end{thebibliography}
\end{document}